\DocumentMetadata{}
\documentclass[acmsmall,screen]{acmart}

\usepackage{url}
\usepackage{pifont}
\usepackage{xspace}
\usepackage{amsthm}
\usepackage{CJKutf8}
\usepackage{hyperref}
\usepackage{makecell}
\usepackage{multirow}
\usepackage{graphicx}
\usepackage{listings}
\usepackage{algorithm}
\usepackage{enumerate}
\usepackage{todonotes}
\usepackage{algpseudocode}
\usepackage{tablefootnote}
\usepackage{color, xcolor}
\usepackage{amsmath, amsfonts}

\newcommand{\vvv}[1]{#1}

\definecolor{dkgreen}{rgb}{0, 0.6, 0}
\definecolor{gray}{rgb}{0.3, 0.3, 0.3}
\definecolor{bg}{rgb}{0.97, 0.97, 0.97}
\definecolor{mauve}{rgb}{0.58, 0, 0.82}
\newcommand{\ie}{\textit{i.e.,}\xspace}
\newcommand{\eg}{\textit{e.g.,}\xspace}
\newcommand{\etc}{\textit{etc.}\xspace}
\newcommand{\etal}{\textit{et al.}\xspace}

\newcommand{\tabref}[1]{Table~\ref{#1}\xspace}
\newcommand{\figref}[1]{Fig.~\ref{#1}\xspace}
\newcommand{\secref}[1]{Section~\ref{#1}\xspace}
\newcommand{\forref}[1]{Formula~\ref{#1}\xspace}
\newcommand{\algopref}[1]{Algorithm~\ref{#1}\xspace}
\newcommand{\coderef}[1]{Code Snippet~\ref{#1}\xspace}

\newcommand{\id}{\textit{ID}\xspace}
\newcommand{\ids}{\textit{ID}s\xspace}
\newcommand{\xpath}{\textit{XPath}\xspace}
\newcommand{\xpaths}{\textit{XPath}s\xspace}
\newcommand{\gui}{{\sc GUIIntention}\xspace}
\newcommand{\code}{{\sc CodeIntention}\xspace}
\newcommand{\toolname}{{\sc TestIntention}\xspace}
\newcommand{\cdsq}{{code2seq}\xspace}
\newcommand{\dome}{{DOME}\xspace}
\newcommand{\llm}{{LLM-basis}\xspace}

\setcopyright{acmcopyright}
\copyrightyear{2023}
\acmYear{2023}

\acmJournal{TOSEM}
\acmVolume{0}
\acmNumber{0}
\acmArticle{1}
\acmMonth{0}

\begin{document}

\title{Test Script Intention Generation for Mobile Application via GUI Image and Code Understanding}

\author{Shengcheng Yu}
\email{yusc@smail.nju.edu.cn}
\orcid{0000-0003-4640-8637}
\affiliation{
  \institution{State Key Laboratory for Novel Software Technology, Nanjing University}
  \city{Nanjing}
  \country{China}
  \postcode{210093}
}
\author{Chunrong Fang}
\authornote{Chunrong Fang and Jia Liu are the corresponding authors.}
\email{fangchunrong@nju.edu.cn}
\orcid{0000-0002-9930-7111}
\affiliation{
  \institution{State Key Laboratory for Novel Software Technology, Nanjing University}
  \city{Nanjing}
  \country{China}
  \postcode{210093}
}
\author{Jia Liu}
\authornotemark[1]
\email{liujia@nju.edu.cn}
\orcid{0000-0001-8368-4898}
\affiliation{
  \institution{State Key Laboratory for Novel Software Technology, Nanjing University}
  \city{Nanjing}
  \country{China}
  \postcode{210093}
}
\author{Zhenyu Chen}
\email{zychen@nju.edu.cn}
\orcid{0000-0002-9592-7022}
\affiliation{
  \institution{State Key Laboratory for Novel Software Technology, Nanjing University}
  \city{Nanjing}
  \country{China}
  \postcode{210093}
}

\begin{abstract}

Testing is the most direct and effective technique to ensure software quality. Test scripts always play a more important role in mobile app testing than test cases for source code, due to the GUI-intensive and event-driven characteristics of mobile applications (app). Test scripts focus on user interactions and the corresponding response events, which is significant for testing the target app functionalities. Therefore, it is critical to understand the test scripts for better script maintenance and modification. There exist some mature code understanding (\ie code comment generation, code summarization) technologies that can be directly applied to functionality source code with business logic. However, such technologies will have difficulties when being applied to test scripts, because test scripts are loosely linked to apps under test (AUT) by widget selectors, and do not contain business logic themselves.

In order to solve the test script understanding gap, this paper presents a novel approach, namely \toolname, to infer the intention of GUI test scripts. \textbf{Test intention refers to the user expectations of app behaviors for specific operations.} \toolname formalizes test scripts with an operation sequence model. For each operation within the sequence, \toolname extracts the target widget selector and links the selector to the GUI layout information or the corresponding response events. For widgets identified by \xpath, \toolname utilizes the image understanding technologies to explore the detailed information of the widget images, the intention of which is understood with a deep learning model. For widgets identified by \id, \toolname first maps the selectors to the response methods with business logic, and then adopts code understanding technologies to describe code in natural language form. Results of all operations are combined to generate test intention for test scripts. An empirical experiment including different metrics proves the outstanding performance of \toolname, outperforming baselines by much. Also, it is shown that \toolname can save about \vvv{80\%} developers' time to understand test scripts.

\end{abstract}

\ccsdesc[500]{Software and its engineering~Software testing and debugging}

\keywords{Mobile App Testing, GUI Understanding, Code Understanding}

\maketitle

\section{Introduction}

With the rapid development of the mobile internet and smart devices, mobile apps have had a pivotal position in a large number of aspects of the whole society and common people's daily lives. Therefore, app quality assurance has become an important topic for both academia and industry. During app development, testing is the most widely-used and most effective way for app quality assurance \cite{harman2016mobile}. As an effective black-box testing approach (relying on the app \textit{apk} files and without obtaining source code), developing GUI (graphical user interface) test scripts based on automated frameworks or techniques, \ie Appium, is a practical approach for the mobile app testing scenario. GUI test scripts can faithfully simulate human user events and are reusable in regression testing. Developers do not have to develop brand new test scripts after app version iteration for original functionalities.

During mobile app maintenance, developers spend a large amount of time understanding existing GUI test scripts. According to Xia \etal \cite{xia2017measuring}, 59\% of time is spent on programming comprehension activities. A study by Sridhara \etal \cite{sridhara2010towards} shows that good comments can greatly improve comprehension efficiency. Despite the significance of code comments, most existing or under-development test scripts are poorly commented on. It is difficult to require developers to write unified easy-to-understand comments manually, and it is even harder to write appropriate comments for large-scale existing test scripts. Successors can hardly fully understand what functionalities are tested in the test scripts with the iteration of the developers in command of the test scripts.

Therefore, automated approaches are imperative to assist in understanding the \textbf{target functionalities and test goals of test scripts}, which can be viewed as \textbf{user expectations of app behaviors for specific testing operations in test scripts}, which we refer to as \textbf{\textit{test script intentions}} in this paper. We provide examples for \textbf{\textit{test script intentions}} in our online supplementary package.

There exist many studies focusing on software source code understanding \cite{haije2016automatic, hu2018deep, hu2020deep, liang2018automatic, song2019survey, wong2013autocomment}, which use information retrieval-based or learning-based technologies to automatically generate comments for source code. However, there is no study for the test script code understanding. Besides, it is not applicable to apply the source code understanding studies to test script code understanding, because app source code and test script code are quite different. App source code are logically self-contained, and express business logic themselves. However, test script code are the target widget location by \id or \xpath and widget operation. The business logic is not fully contained in the test script code, and we have to refer to other information to obtain the business logic for code understanding, \eg related response methods and GUI information. 

In this paper, we propose a novel approach, \toolname, to infer the test script intentions and generate natural language descriptions. One crucial prerequisite of test script intention understanding is the formal representation of test scripts. For each test script, we model it as an operation sequence. Each operation sequence is composed of a sequence of operations, like click actions, input actions, long-press actions \etc In \toolname, we collect the test intentions of each operation separately to acquire the operation intention. Afterward, we aggregate the test intentions of all the operations and get the intention of the whole test script.

Considering the GUI-intensive and event-driven features of mobile apps, information from both app GUI and the underlying source code containing business logic are important for the test script intention obtaining. Therefore, \toolname is composed of two parts, \gui and \code, which analyze the test script and the AUT from the GUI level and the source code level, respectively. 

\gui includes textual information retrieval and widget image understanding. Mobile apps are GUI-intensive software, so much information can be obtained from GUI images with computer vision (CV) technologies, including textual information and non-textual information. For the formalized test scripts, \toolname captures the GUI screenshot of each operation, and the screenshots are processed with the perspective of rich information collections of widgets instead of simple images consisting of pixels \cite{yu2021layout}. CV algorithms are applied to extract all the widgets from the screenshots. Within the widgets, textual information, which is supposed to reflect the widget intentions, can be recognized with the OCR algorithm and should be merged according to the GUI layout files. Non-textual information cannot be directly obtained, so we can infer the widget intentions by a pre-trained DL model, which is trained by a large-scale self-constructed widget image dataset. 

\code contains the response method mapping and code intention generation. With the widget selector (\id), the response method snippets are expected. However, the mapping process from widget selector to response method is challenging due to the dynamic binding of \ids and the explicit binding mechanism of the Android platform. We design a novel algorithm adopting static analysis to the AUT, with the goal of locating the response method snippets from the app source code. With such one or several response code snippets, \toolname adopts the code understanding technologies to infer the code intentions. Specifically, the code snippets extracted from the static analysis to the AUT are fed into a pre-trained RNN model, which extracts code features and generates code intention descriptions in a natural language format.

\toolname merges the results from \gui and \code and conducts text deduplication. Till then, \toolname can effectively generate test intentions for the given GUI test scripts. We further design and implement a tool to evaluate the effectiveness of \toolname. Results show that \toolname outperforms the traditional code understanding approach by much on different metrics, and can effectively generate test script intentions for the reference of app developers. A user study also shows that \toolname can significantly save about \vvv{80\%} of app developers' time to review and understand the test script intentions.

In this paper, we declare the following noteworthy contributions:

\begin{itemize}
	\item We propose a novel approach, \toolname, utilizing GUI image and code understanding to fully explore the code, textual, and image information, and then \toolname generates test script intentions in natural language.
	\item We present a novel algorithm to map test operations in test scripts to the corresponding response methods in source code and GUI image information in the GUI layout files.
	\item We propose a test operation sequence model to formally represent the test script, and include the necessary information to map the operations, like operation type, selector, operation, and specific contents.
	\item An empirical experiment and a user study show the effectiveness of \toolname and the capability to alleviate developers' burden to understand test scripts.
\end{itemize}

\textbf{A supplementary package can be found on \url{https://sites.google.com/view/testintention}.}

\section{Challenge and Motivation}

\subsection{Script Intention Understanding}
\label{sec:siu}

Test scripts are poorly commented on, making it challenging for developers to understand test script intentions. During the maintenance or modification of test scripts, developers first obtain the test script intentions via GUI transitions. Then, in order to precisely know which operation triggers which GUI transition, developers have to manually locate the target widgets with the selectors\footnote{\url{https://www.w3.org/TR/webdriver/\#locator-strategies}} (\id or \xpath) for each operation step by step. \vvv{There are several different types of selectors, including CSS selector, link text selector, partial link text selector, tag name, and XPath selector. Such selectors are all supported by the most popular and widely used mobile app GUI test script development framework, Appium\footnote{\url{http://appium.io/docs/en/latest/}}. According to our investigation (details described below), \id and \xpath are the most common used selectors in real-world industrial practice, which takes a percentage of 53.8\% and 45.7\%, respectively. Therefore, we mainly research on dealing with these two types. As the characteristics of the architecture of a mobile app is that the front-page information stored in GUI layout files and the back-end code business logic are in different layers and relatively isolated, the response methods are associated with GUI widgets with such selectors. The selectors are either statically coded in the app GUI layout files or dynamically generated and allocated during runtime.} It is a hard and time-consuming task to locate the target widgets one by one and assemble the intentions of the whole test script.  

To illustrate the severity of this problem, we have investigated the comment situation of GUI test scripts. We collect a dataset of \vvv{200} test scripts for \vvv{10} different mobile apps. The test scripts are developed by \vvv{132} different testers. The testers are skillful test engineers with at least two years of mobile app testing experience. With the collected test scripts, we require the two mobile app testing engineers to estimate the comprehensibility of the test scripts according to the amount and completeness of source code comments. \vvv{The apps and test scripts are all from real-world industrial companies. We collect such apps and GUI test scripts from our collaborators, who directly grant access to us to form the dataset. Such apps cover different categories, like Communication, Tool, Finance, \etc The 10 mobile apps are all from different companies, including emerging software companies and traditional industrial companies. Such factors help improve the generalizability of our investigation. The results show that 64.6\% of the test scripts have no comments to illustrate the test script intentions or operation intentions. 26.7\% of the test scripts have few comments that simply help the test script developers themselves to understand the intentions, while the comments are hard for others to understand. Only 8.7\% of the test scripts are with complete comments.}

We also define a metric named ``comment-code ratio'' to further illustrate such a problem, which is the ratio of the line of comments and line of code, and can be expressed as $ratio = \frac{LOC_{comment}}{LOC_{code}}$. If $ratio$ is higher than 0.3, the test is considered well commented. If $ratio$ is between (0, 0.3], the test is considered commented. We conduct a manual investigation on 100 randomly-selected apps from an open dataset AndroZooOpen\footnote{\url{https://github.com/SMAT-Lab/AndroZooOpen}} \cite{liu2020androzooopen}. The results show that only 1\% of tests are well commented. For the rest, 33\% have some fragmentary comments showing the test intentions. That is to say, most (66\%) of the tests have no comments, leading to an obstacle to test understanding. This survey and the judgment of the completeness of the test code are conducted and verified by two senior software testing engineers. The results show that a large majority of app developers are not used to writing comments for the test part of the project.

As presented in existing research designed to automatically generate code comments for source code with business logic, developers hardly write complete and generally comprehensible code comments \cite{haije2016automatic, hu2018deep, hu2020deep, liang2018automatic, song2019survey, wong2013autocomment}. Such approaches behave well on code comment generation tasks as shown in corresponding papers, but situations are different for GUI test scripts of mobile apps. For GUI test scripts, the business logic heavily depends on AUT instead of being self-harmonious. To fully and appropriately generate test script intentions, it is necessary to map test operations to corresponding response information. In this paper, we try to simulate and automate the manual process to understand the test script intentions. For each operation in the operation sequence, we need to locate the GUI information, or response methods, with the selectors, \ie \ids or \xpaths, which is helpful to infer the test operation intentions. Another problem is how to precisely extract effective response methods and exclude the distractions after localization, so we need to conduct the code slicing.

\subsection{A Motivating Example}

In this section, we provide an example to illustrate the motivation of this work. \coderef{code:script} presents part of a test script for the app \textit{ContactManager} (used in our experiment), including the ground-truth intentions and the intentions generated by \toolname. \vvv{We also present the GUI transitions of the app with the execution of \coderef{code:script} to better present the intention of the test script.} In this example, \vvv{Line 7--22} represents operations ($Op$ defined in \secref{sec:opseq}) and corresponding intentions and forms the operation sequence ($OpSeq$ defined in \secref{sec:opseq}) with six $Op$s.

\begin{figure}[!h]
   \centering
   \includegraphics[width=\linewidth]{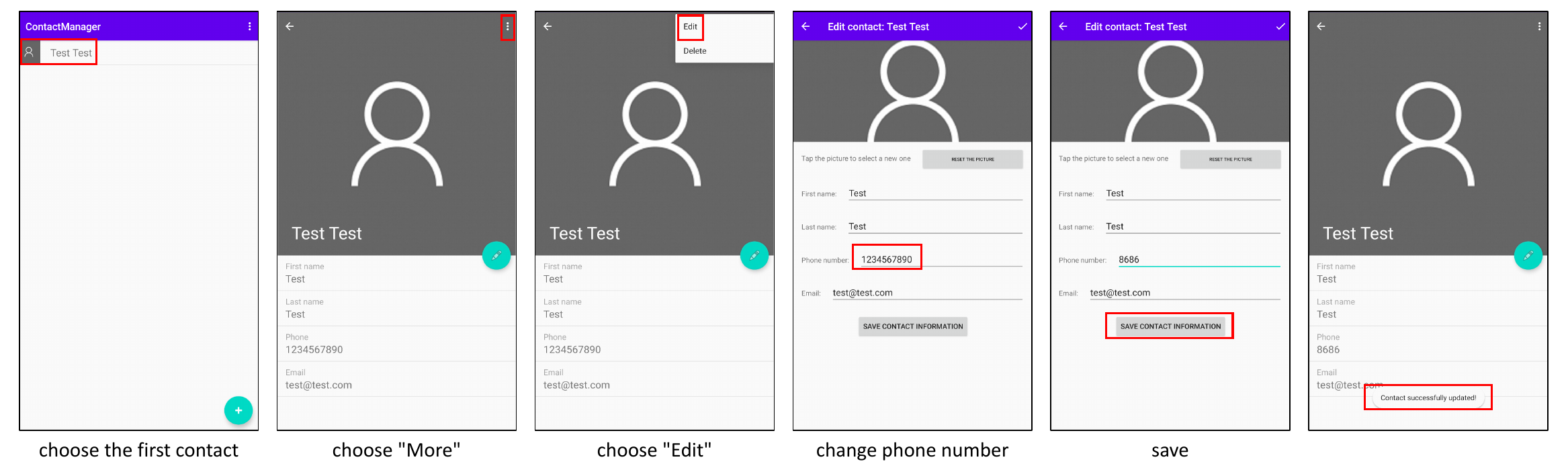}
   \caption{\vvv{GUI Transitions with \coderef{code:script}}}
   \label{fig:example}
\end{figure}

In the $OpSeq$, the target widgets are located by two categories of widget selectors, the \ids (\vvv{Line 18, Line 19, Line 22}) and the \xpaths (\vvv{Line 9, Line 12, Line 15}). In the test script, the only information we can obtain is the widget selector and the corresponding operation types (\eg \texttt{click()}, \texttt{sendKeys()}). Due to the lack of the presence of the logic business in test scripts, it is hard for automated tools or learning-based models to infer the intentions of each operation, and the whole operation sequence. 

We hold it is a significant part to map the test operations to the app source code with the business logic or any information from the GUI layout files. However, existing approaches focus on the code that directly reflects functionalities or business logic.

\begin{lstlisting}[
	language = Java,
	caption  = Example for Mobile App Test Script,
	label    = code:script
]
public class ContactManagerTest {
  // ground-truth:  edit the contact
  // TestIntention: contact edit and save submit
  public void editContact(AppiumDriver driver) {
    //initialization work
   ...
    // ground-truth:  choose the first contact
    // TestIntention: select first contact
    driver.findElementByXPath("/hierarchy/FrameLayout/.../LinearLayout[1]").click();
    // ground-truth:  choose "More"
    // TestIntention: click more button
    driver.findElementByXPath("//ImageView[@content-desc=\"More\"]").click();
    // ground-truth:  choose "Edit"
    // TestIntention: click edit button
    driver.findElementByXPath("/hierarchy/FrameLayout/.../Layout[1]").click();
    // ground-truth:  change phone number
    // TestIntention: input phone number
    driver.findElementById("com.m.g.c:id/inputphone").clear();
    driver.findElementById("com.m.g.c:id/inputphone").sendKeys("8686");
    // ground-truth:  save
    // TestIntention: save submit
    driver.findElementById("com.m.g.c:id/actionsubmit").click();
  }

  public static void main(String[] args) {
    // other configurations
    ...
  }
}
\end{lstlisting}

\subsection{Response Method Mapping}
\label{sec:mapping}

Response method refers to a \textit{Listener} method that responds to a specific action on a specific widget. The accurate mapping of the response method from test scripts to the business logic in source code is a spiny problem. According to our investigation, all the widgets in the test scripts are located by a unique selector, which is an \id or an \xpath. For widgets located by \ids, the response methods can be located because the methods are bound to the widgets explicitly with the \id. However, for widgets located by the \xpath, the response method mapping is more complicated. Due to the event-driven feature, some widgets are dynamically generated, and the response methods are also dynamically bound to the widgets. Therefore, implicit underlying mechanisms of Android complete the binding, hindering manually locating corresponding response methods. 

According to our study on the test scripts, we find that due to not strictly following the coding criteria, more than 50\% widgets have to be located by \xpath in the test scripts, lacking the \id labeled. Among the widgets located by \xpath, about 77\% are located with the hierarchy, and the rest are located with the XML attribute \texttt{content-desc}, which shows the visible texts of the widgets.

\begin{lstlisting}[
	language = Java,
	caption  = Example for Implicit Response Method Binding,
	label    = code:example
]
  mAdapter = new CityAdapter(this, mList);
  mAdapter.setOnItemClickListener(new CityAdapter.OnItemClickListener() {
      @Override
      public void onItemClick(View v, ImageView favo, int position) {
          LocationEntity entity = mList.get(position);
          final FavoriteEntity favoriteEntity = new FavoriteEntity(entity.getId(), entity.getName(), entity.getPath());
          mViewModel.insertFavorite(favoriteEntity);
          entity.setFavorite(true);
          favo.setSelected(true);
      }
  });
  recyclerView.setLayoutManager(new LinearLayoutManager(this));
  recyclerView.setAdapter(mAdapter);
\end{lstlisting}

One vivid example of implicit response method binding is the \textbf{Adapter}. the example is shown in \coderef{code:example}. This operation is from an app named \textit{iWeather}. Users click on the search box and input the city name, and the app will post a request and return the data. The returned data is adapted with the \texttt{CityAdapter}. The \texttt{recyclerView} utilizes the adapter to render the data to the GUI. Such a process is dynamic, and the list items are dynamically generated, so it is hard to bind the response method statically.

\section{Approach}

\begin{figure}[!h]
   \centering
   \includegraphics[width=\linewidth]{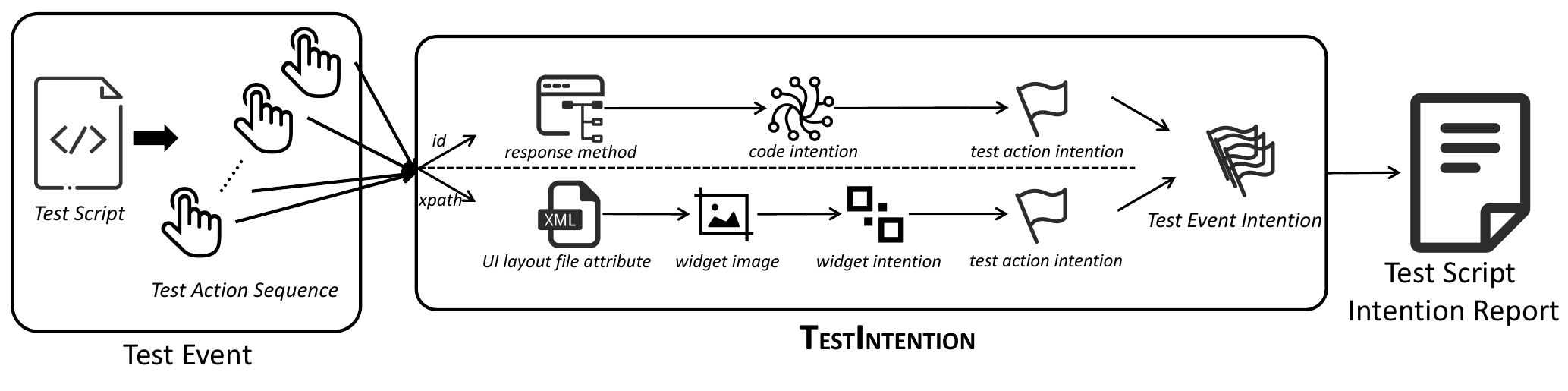}
   \caption{\toolname Framework}
   \label{fig:framework}
\end{figure}

In this section, we illustrate the general framework of \toolname. \toolname is composed of two parts, the \gui, and the \code. The results from the two parts are merged to get the final test script intentions. \gui, targeted at the widgets located by \xpath, searches for the information from GUI layout files and GUI images. For widgets located by \id, corresponding response methods are obtained and analyzed in \code to infer the intentions. \figref{fig:framework} presents the general framework of \toolname. 

\subsection{Operation Sequence Model}
\label{sec:opseq}

Test scripts are unstructured textual information. Some driver initialization or post-processing code should not be considered in the test intention generation. The relevant information in test scripts to script intention is the operations. For the mobile app test scripts, each operation ($Op$) in the operation sequence ($OpSeq$) contains an element selector and a concrete operation. In order to extract the critical information linked to the test operations and eliminate irrelevant information in the test scripts, we use regular expression matching to extract all operations and formally express the test script with the operation sequence model. Irrelevant code in GUI test scripts will be left out by the regular expression matching, and only meaningful testing operations will be extracted. Then, the operations from test scripts are formed with the operation sequence model as follows:

\begin{equation}
OpSeq = \{Op_1, Op_2, Op_3 ... Op_i ... Op_n\}
\end{equation}

where the $Op_i$ is a 4-dimension tuple:

\begin{equation}
Op_i = \{type, selector, operation, content\}
\end{equation}

The $type$ determines whether the widget is located by \id or by \xpath, which are two main widget selector categories as illustrated in \secref{sec:mapping} Different kinds of selectors are processed with different technologies. The $selector$ is the concrete locating selectors of the widget, \ie ``\texttt{com.app.calendar:id/switcher\_layout}'', ``\texttt{/hierarchy/LinearLayout[2]/Button[2]}''. The $operation$ refers to the concrete operation, \ie \texttt{click()}, \texttt{sendKeys()}. The $content$ is a parameter of some specific $operation$s. For example, the \texttt{sendKeys()} method should be assigned with a string, like for the username field, the method should be invoked as \texttt{sendKeys(``admin'')}, where the ``admin'' is the $content$ in this statement. The $content$ is not applicable for all types of operations, \eg \texttt{click()}, and only some of the $operation$s should be assigned with some $content$s. 

\vvv{Based on the operation sequence model, we define the concept of \textbf{intention}. For each $Op_i$ in $OpSeq$, there is a corresponding \textbf{intention}, which we name it as the \textbf{operation intention}, referring to the user expectations of app behaviors for specific operations:}

\begin{equation}
Intention_{Op_i} \leftarrow \left\{ Op_i | be \leftarrow ue \right\}
\end{equation}

\vvv{where $Intention$ refers to the test intention, $Op_i$ refers to the target operation, $be$ refers to the app behaviors, and $ue$ refers to the corresponding user expectations. As $Op_i$ composes the $OpSeq$, which is the formal representative of the GUI test script, \textbf{operation intentions} of different operations form the \textbf{intention} of the whole GUI test script:}

\begin{equation}
Intention_{OpSeq} = merge\{Intention_{Op_1}, Intention_{Op_2}, ... Intention_{Op_i} ... Intention_{Op_n}\}
\end{equation}

\vvv{In our approach, \gui and \code are two parts that are designed to deal with $Op_i$ that may contain different $selector$s, \ie \xpath or \id. Examples can be seen from \coderef{code:script}.}

\subsection{\gui}
\label{sec:gui}

\vvv{\gui is mainly targeted at the widgets located by \xpath. Such widgets are hard to be mapped to the response methods because such widgets are implicitly linked with response methods as illustrated in \secref{sec:mapping}. However, these widgets are more closely linked to the GUI layout files. Therefore, \toolname uses \gui to obtain information and generate script intentions from the GUI level. Intuitively, \xpaths are formed by the hierarchy structure, which can be utilized to locate the target widgets from the GUI layout files. Such files contain rich textual information, which indicates the intentions of corresponding widgets. Besides, mobile apps are GUI-intensive, so we can infer the widget intentions from the widget screenshots with image understanding technologies.}

Widgets located by \xpaths are divided into two categories\footnote{\url{https://appium.io/docs/en/commands/element/find-elements/}}. The first category is in the format of \forref{equ:content}. For such a type, we use regular expression matching to match the \texttt{widgetType} field and the \texttt{content-desc} field.

\begin{equation}
	\label{equ:content}
	\texttt{//<widgetType>[@content-desc=``<text>'']}
\end{equation}

\begin{equation}
	\label{equ:hierarchy}
	\begin{aligned}
	\texttt{/hierarchy/<view>/../<view>[i]/../<view>}	
	\end{aligned}
\end{equation}

The second type of widgets located by \xpath is on the basis of XML hierarchy structure (\forref{equ:hierarchy}), and \toolname splits the \xpath by slashes. Widgets without index indicate that such a widget type can be uniquely determined under the former widget. Widgets with index indicate that such widgets cannot be uniquely determined by the type, so numerical order is needed. Widgets are located from the root node of \xpath to search for the GUI layout files. With the located widgets, \gui uses Textual Information Retrieval and Widget Image Understanding to infer the test intentions.

\vvv{\toolname adaptively retrieves textual information from the app. For the widgets that are attached with textual information from GUI layout files, like \texttt{content-desc}, \toolname directly extracts such texts. However, sometimes the textual information cannot be obtained from GUI layout files if they are nested in the \texttt{Canvas} widget or HTML pages. For such widgets, \toolname uses OCR algorithms to extract texts directly from app screenshots (\secref{sec:tir}). It is more often that GUI widgets are not explicitly linked with textual information, and under such circumstances, we use the image captioning model to identify the intentions of widget images (\secref{sec:wiu}).}

\subsubsection{\textbf{Textual Information Retrieval}}
\label{sec:tir}

For current mobile apps, textual information extracted from the GUI is rich, and such textual information can provide accurate information about the widget intentions. Typical examples are easy to find like the ``Submit'' button, ``Cancel'' button, ``Username'' text box, ``Password'' text box, \etc Therefore, we hold that such textual information is valuable for us to infer the test script intentions. 

However, as discussed above, the textual information of some widgets cannot be obtained from the original GUI layout files because some widgets are dynamically generated. Therefore, \toolname has to dump the GUI layout files at runtime. UIAutomator\footnote{\url{https://developer.android.com/training/testing/ui-automator}} can be used to manually dump the runtime GUI layout files, while in order to automate such a process, we design a driver wrapper and add the functionality to obtain the information, including:

\begin{itemize}
	\item \underline{\textbf{App screenshot}} can be utilized to extract existing textual information from the app activities of the mobile apps.
	\item \underline{\textbf{Widget screenshot}} may have implied information about the widget intentions, which cannot be directly extracted from screenshots.
	\item \underline{\textbf{Attribute textual information}} is contained in the GUI layout files and is not shown on GUI (\ie widget name, content description, \etc), which can help infer and generate widget intentions.
\end{itemize}

The process of the app screenshot and the widget screenshot will be presented in detail in \secref{sec:wiu}. The widget elements have many different attributes. After the target widget is determined, we extract the intention-related attributes, including \texttt{text} and \texttt{content-desc}. However, there is some textual information that cannot be acquired from the runtime GUI layout files. We utilize CV technologies to further process the app screenshots, and the specific technology we adopt is optical character recognition (OCR) technologies, which can not only recognize the texts on the screenshot but also is capable of locating the text coordinates. From the runtime GUI layout files, we can also acquire the location of the target widget. We match the OCR results and the widget coordinates from the runtime GUI layout files and obtain the GUI-level textual information. Finally, the textual information both from runtime GUI layout files and GUI images is combined to generate the intentions of the target widgets.

\subsubsection{\textbf{Widget Image Understanding}}
\label{sec:wiu}

\begin{figure}[!h]
\centering
\includegraphics[width=0.5\linewidth]{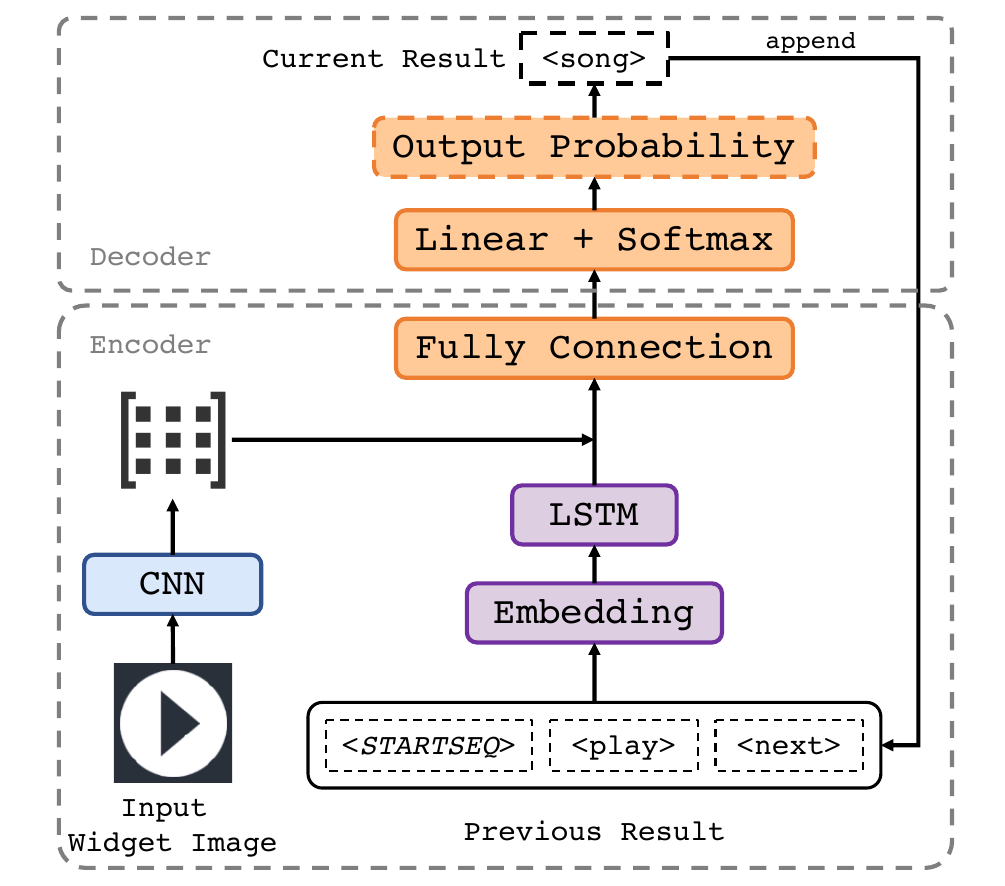}
\caption{Widget Intention Generation Model}
\label{fig:widget}
\end{figure}

Some widgets attached with explicit textual information can be processed with GUI layout files or OCR technologies on app screenshots. However, circumstances are more often that the majority of the widgets are not attached with any explicit textual information even in layout files, \eg a magnifier icon referring to the ``search'' functionality, three points referring to the ``more'' menu, \etc Such widgets do not have any textual information on app screenshots or in textual attributes of GUI layout files. To better understand the widgets without textual information and generate captions for the widget images, we construct a deep learning model with an ``Encoder-Decoder'' structure on the basis of \cite{chen2020unblind} (\figref{fig:widget}).

The encoder part encodes the input image and the caption texts, respectively, including the widget feature extractor and the sequence feature processor. 

The widget feature extractor is designed to extract the visual features from the input widget images. We use a convolutional neural network (CNN) to extract the image features, which is widely used to process images and extract features \cite{chen2020unblind}. CNN models can effectively encode image features. The CNN model contains two different types of layers, the convolutional layer, and the pooling layer. For a common CNN model used to execute a classification task, the output is a vector of probabilities that the image belongs to each preset type. However, our goal is to extract image features for further processing, so we remove the last layer and output the feature vector, which is a 256-dimension vector that can effectively represent each input widget image \cite{li2019boosting, jang2020approach}.

The sequence feature processor extracts the features from text input. The initial sequence is a start token, $\langle STARTSEQ \rangle$. The sequence is fed into an embedding layer to encode the natural language into vectors. The advantages are that 1) the generated feature vectors are more concentrated, which avoids the waste of computation resources; 2) it considers the semantics of texts. The embedded words are fed into an LSTM model \cite{hochreiter1997long} to parse the embedded tests into a 256-dimension vector.

After obtaining the two feature vectors of widgets images and current sequences, we concatenate the vectors and get a 512-dimension feature vector. The feature vector is fed into the fully connected layers to further decode the feature. The intermediate result vector is fed into a \texttt{SoftMax} layer to predict the next word of the widget image caption. The output of the \texttt{SoftMax} layer is an output probability vector and can be mapped to a single word in the corpus, which is the current result in \figref{fig:widget}. The current result will be concatenated with the previous result sequence, and the prediction process will be repeated. The end flag is the $\langle ENDSEQ \rangle$.

\begin{figure}[!h]
\centering
\includegraphics[width=\linewidth]{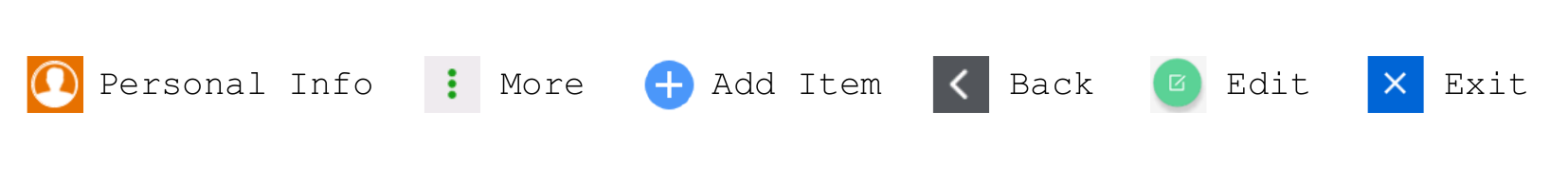}
\caption{Examples of Widget without Texts (the attached texts mean the widget functionalities, \ie intentions.)}
\label{fig:widgetExample}
\end{figure}

The widget image understanding model is composed of two encoders and a decoder. The encoder for widget images we use is the VGG-16 model \cite{simonyan2014very}, which is a representative CNN model and is with low time overhead in widget image encoding. This model is introduced for other tasks \cite{simonyan2014very}, and we fine-tune it to fit our task and scenario. The fine-tuning process is based on a self-built dataset containing 2,000 different widget images, each of which is attached with three pieces of intention captions. We recruit three graduate students to label and cross-validate the captions. Some examples can be seen in \figref{fig:widgetExample}. \vvv{The widgets images are collected from the widely used Rico dataset\footnote{http://www.interactionmining.org/rico.html}, and we manually confirm that the apps in the dataset have no overlap with the apps we use in the model effectiveness evaluation and the \toolname evaluation.} The dataset is divided according to the conventional ratio, 7:2:1, into the training set, the validation set, and the test set. The input image size is $(244, 244, 3)$ after resize operation. The model is composed of a series of convolutional layers, pooling layers, and fully-connected layers. The last layer of the original model is removed, and the last but one layer is taken as the output to effectively represent the widget feature \cite{li2019boosting, jang2020approach}.

\subsection{\code}
\label{sec:code}

Another main locating method of widgets is through \id. Such widgets are closely linked to specific response methods in app source code. Instead, the \ids may be missing in the GUI layout files. Therefore, \toolname adopts \code to infer the intentions of such widgets. For widgets located by \id, we can find the binding response methods and infer the corresponding intentions in the form of natural language by feeding code snippets of the response methods into an RNN model, which is good at processing text fragments of variable length and extract features. \code consists of Response Method Mapping and Code Intention Generation.

\subsubsection{\textbf{Response Method Mapping}}

For operations located by \id, \code has a global search and matching from app source code for the target \ids. After obtaining the source code files containing the target \ids, \toolname uses the template matching to execute the mapping of the response method \vvv{(\ie event handler/listener)}. Five different templates are summarized that are supposed to be able to cover the majority of response methods in real-world Android apps. \vvv{First, we refer to the official Android development documents, including Android API, which provide all the supported response method binding. Such documents provide official guidance to developers on how to develop code as the response methods for different kinds of widgets. Based on the official documents, we summarized five different response method mapping templates, which cover all the supported response method binding ways given in the documents. Second, we refer to actual apps from the AndroZooOpen dataset to check the generalizability of the summarized templates. Two of the authors review the source code of 50 different apps (no overlap with the apps used in the following evaluation), which are with high popularity and representative. Such apps are reviewed to check whether there are any extra response method binding ways that are not depicted in the official Android development documents, and the review results show that all the actual practice are included in our summarized templates.}

\begin{lstlisting}[
	language = Java, 
	caption  = Example for Switch Statement Template, 
	label    = code:template1
]
  case R.id.action_pin_recipe_to_widget:
	  pinRecipeToAppWidget();
	  return true;
\end{lstlisting}

\ding{202} The first template is in the form of the switch statements. An example is in \coderef{code:template1}. In such a template, the \id is used as the condition to trigger different response methods.

\begin{lstlisting}[
	language = Java, 
	caption  = Example for If-Else Statement Template, 
	label    = code:template2
]
  if(item.getItemId() == R.id.action_settings) {
	  startActivity(new Intent(this, SettingsActivity.class));
  }
\end{lstlisting}

\ding{203} The second template is in the form of if-else statements (\coderef{code:template2}). Similar to the first one, the \id is used as the judging condition to trigger different response methods.

\begin{lstlisting}[
	language = Java, 
	caption  = Example for ID Binding Template, 
	label    = code:template3
]
  // 1
  TextView tv = findElementsById(R.id.xxx);

  // 2
  @BindView(R.id.rv_steps)
  RecyclerView stepsRecyclerView;
\end{lstlisting}

\ding{204} The third template is in the form of ID binding (\coderef{code:template3}). Specifically, the \id is bound to a widget object. There are two sub-types. For the first sub-type, the \id acts as a parameter of the \texttt{findElementById()} method. For the second sub-type, the \id acts as a parameter of a \texttt{@BindView} annotation label and is assigned to a member variable.

\begin{lstlisting}[
	language = Java, 
	caption  = Example for \texttt{@OnClick} Annotation Template, 
	label    = code:template4
]
  @OnClick(R.id.btn_next_step)
  void openNextStep(){
    	//...
  }
\end{lstlisting}

\ding{205} The fourth template is in the form of \texttt{@OnClick} annotation (\coderef{code:template4}). The \id is a parameter for the \texttt{@OnClick} annotation label, and the annotation label is assigned to a response method, which corresponds to the widget.

\begin{lstlisting}[
	language = Java, 
	caption  = Example for Layout File Attribute Declaration Template, 
	label    = code:template5
]
  < Button
          android:id="@+id/nextButton"
          android:onClick="onNext"
          android:text="@string/next"
  />
\end{lstlisting}

\ding{206} The last template is in the form of the layout file attribute declaration (\coderef{code:template5}). This template is the only one that is matched by the GUI layout files. In the attributes of the widget, the \texttt{android:onClick} attribute explicitly shows the response methods.

\begin{lstlisting}[
	language = Java, 
	caption  = Code Template Prioritization, 
	label    = code:prior
]
  @OnClick({R.id.fab_search, R.id.fab_region})
  public void onClick(View view){
      fam.collapse();
      if (view.getId() == R.id.fab_search) {
          Intention intention = new Intention(this, SearchActivity.class);
          startActivity(intention);
      } else if (view.getId() == R.id.fab_region) {
          Intention intention = new Intention(this, RegionActivity.class);
          startActivity(intention);
      }
  }
\end{lstlisting}

For the aforementioned five templates, we rank them according to their priority (from \ding{202} to \ding{206}), and the priority is determined by their appearing frequency according to our survey mentioned in \secref{sec:mapping}. Such a priority helps reduce the time required for the response method search and template matching. For example, as shown in \coderef{code:prior} for the app named \textit{iWeather}, the response method can be located both by template \ding{203} and by template \ding{205}. However, if we use template \ding{205}, we need a further process to eliminate the influence of another \id. If we use template \ding{203}, another \id will not be considered.

After \toolname obtains the response methods, code intention generation may still encounter problems. For example, in \coderef{code:template1}, the widget \id is bound to an encapsulated method. This can lead to information insufficiency and invalidate the code intention generation algorithm. To alleviate the negative effect, we introduce a nested search to complement the information. Specifically, we locate the method body and replace the method name within the located response method body (no iterative search is conducted because the information can be inferred with the one-level nesting). 

\subsubsection{\textbf{Code Intention Generation}}

We use another ``Encoder-Decoder'' structure deep learning model to understand the code snippet of the response method and generate the code intentions. Our model is based on the \cdsq model proposed by Alon \etal \cite{alon2018code2seq}. The \cdsq model is one of the state-of-the-art and representative approaches for the app source code comment generation, and will not bring much time overhead. In natural language processing tasks, the input of the model is the natural language fragments. However, for code snippet input, texts cannot be directly processed because they contain specific semantics and syntax. As used in many program understanding models like \cite{alon2018code2seq, alon2019code2vec}, code snippets are transferred into abstract syntax trees (AST) to keep the semantics and syntax information. AST is a 6-dimension tuple: $AST = \langle N, T, X, s, delta, val \rangle$, where the $N$ is the non-leaf node set; $T$ is the leaf node set; $X$ is the leaf node value set; $s$ is the root node; $delta$ is the map from $N$ to $N \cup T$, which means the relationship of father nodes and the corresponding son nodes, and $val$ is the map from $T$ to $X$, which binds each leaf node with the specific value \cite{allamanis2015suggesting, allamanis2016convolutional}.

\begin{figure}[!h]
\centering
\includegraphics[width=0.5\linewidth]{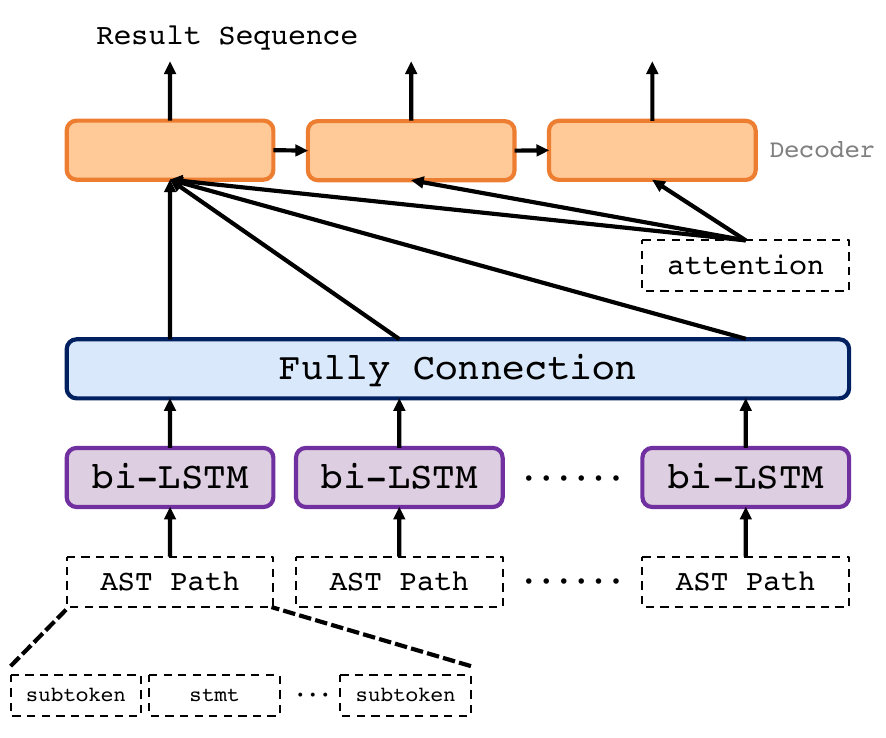}
\caption{Code Intention Generation Model}
\label{fig:code}
\end{figure}

Afterward, the AST of the transferred code snippet (\figref{fig:code}) is fed into the code intention generation model. First, the paths are encoded with a bi-directional LSTM model to a 128-vector, and the path vectors are merged with an average calculation. Then, the average vector is fed into the decoder with an attention mechanism \cite{bahdanau2014neural}. The decoder decodes the extracted AST path vector into a result sequence in natural language, representing the code snippet intention.

The code intention generation model is also in the form of the ``encoder-decoder'' structure. For the model, we use the \texttt{sparse\_softmax} \texttt{\_cross\_entropy} method as the loss function. The optimization method is the \textit{Nesterov momentum} method. We train the model for 3,000 epochs, and the batch size is 512, and for other settings, we follow the original \cdsq model. The corpus size for generating code snippet intention is 27,000. The original \cdsq model is trained with 9,500 Java projects. In order to make the model better fit our scenario, we add 100 more mobile app projects randomly selected from the AndroZooOpen dataset. \vvv{We manually confirm the projects have no overlap with the apps used in the evaluation to avoid the potential data leakage issue.}

\subsection{\vvv{Intention Aggregation}}

\vvv{With the intentions of operations of the operation sequence being acquired, we aggregate them to get the final test intention of the whole operation sequence. In order to merge the similar intentions of different operations, we use a Word2Vec model \cite{mikolov2013distributed} to judge the similarity among different operation intentions, and the similar ones are deduplicated and then the others are concatenated as test script intentions. The output is a highly comprehensible report presenting the test intention of the whole operation sequence and each operation.}

\begin{algorithm}[h]
	\caption{\toolname Workflow}
	\label{alg:testintention}
	\begin{algorithmic}[1]
		\Require Test Script $script$
		\Ensure Test Script Intention Report $report$
		\State $te$ = formalize($script$)
		\State initiate intention list $intentions$
		\For{Test Action $ta \in te$}
			\If{$ta[0]$ == `xpath'}
				\State $widget$ = widgetImage($ta[1]$)
				\State $intention_v$ = widgetImageUnderstanding($widget$)
				\State $intention_t$ = widgetTextExtraction($ta[1]$)
				\State $intention$ = merge($intention_v$, $intention_t$)
				\State $intentions$.add($intention$)
			\Else
				\State // ta[0] == `id'
				\State $response$ = locateResponseMethod($ta[1]$)
				\State $intention$ = codeIntention($response$)
				\State $intentions$.add($intention$)
			\EndIf
		\EndFor
		\State $report$ = generateIntentionReport($intentions$)
		\State return $report$
	\end{algorithmic}
\end{algorithm}

The workflow of the \toolname generation process is formally described in \algopref{alg:testintention}. The input is the Appium-based GUI test script, and the output is a report presenting the test intentions. The first step is to formalize the test script into an operation sequence (Line 1), consisting of a series of operations, and then the process of the following steps on each operation. If the operation is located by \xpath, we invoke the \gui (Line 4--9) including Textual Information Retrieval (Line 7) and Widget Image Understanding (Line 5--6). If the operation is located by \id (Line 11--14), we invoke the \code including Response Method Mapping (Line 12) and Code Intention Generation (Line 13). The final step is to generate the test intention report by merging single operation intentions (Line 17). Deduplication is conducted to make intentions more comprehensible.

\section{Experiment}

\subsection{Experimental Setup}

\subsubsection{\textbf{Research Questions (RQ)}}

We in total set three research questions (RQ) to respectively evaluate the intention generation effectiveness, the operation mapping effectiveness, and the user experience of \toolname.

\begin{itemize}
	\item \textbf{RQ1: (Intention Generation)} How effectively can \toolname generate natural language test intentions?
	\item \textbf{RQ2: (Operation Mapping)} How effectively can \toolname map the operations of the operation sequence to the response code or GUI information?
	\item \textbf{RQ3: (User Experience)} How much can \toolname improve the developers' efficiency to understand scripts?
\end{itemize}

\subsubsection{\textbf{Data Preparation}}
\label{sec:dp}

In this experiment, we use \vvv{50} mobile apps (in \tabref{tab:app}) from the open dataset AndroZooOpen\footnote{\url{https://github.com/SMAT-Lab/AndroZooOpen}} \cite{liu2020androzooopen} \vvv{because we need the source code for operation mapping}. The advantage of this dataset is that the apps are open-sourced. We can obtain the source code for code understanding and then infer the test intentions. AndroZooOpen contains over 45 thousand of open-source Android apps and their metadata, and it is widely used in existing mobile app research. The apps cover different app categories and domains (\eg \textit{Finance}, \textit{Education}, \textit{\vvv{Communication}}, \etc), which can \vvv{indicate} the generalizability of \toolname. We have the following selecting criteria:

\begin{enumerate}[(1)]
	\item Apps should be developed in pure Java language without Kotlin language. Because Kotlin code has different syntax rules, the static analysis algorithm adopted by \toolname focuses on Java language.
	\item App activities should not be video games. Because the computer vision technologies we adopt focus on the layouts that are tables or lists. Video games contain many activities which are specially rendered, and our approach would not fit such a condition.
\end{enumerate}

\begin{table}[!h]
\caption{\vvv{Mobile Apps in the Experiment}}
\label{tab:app}
\centering
\scalebox{0.72}{
\begin{tabular}{cccc|cccc}
\toprule
App               & Category      & LOC     & Project Size & App                  & Category       & LOC   & Project Size \\ \midrule
AgendaOnce        & Productivity  & 0.5k    & 175 KB       & Calculator           & Tool           & 0.2k  & 236 KB       \\
AppIconLoader     & Tool          & 1.5k    & 923 KB       & CapstoneHealth       & Health         & 0.7k  & 271 KB       \\
Baking            & Education     & 1.7k    & 5.1 MB       & CoolWeather          & Weather        & 1.7k  & 1.8 MB       \\
CallBlocker       & Communication & 1.1k    & 433 KB       & FuelUp               & Finance        & 6.8k  & 767 KB       \\
ContactManager    & Business      & 0.8k    & 3.1 MB       & GoalApp              & Productivity   & 1.2k  & 369 KB       \\
CourseAssist      & Education     & 2.8k    & 10.3 MB      & KuanChat             & Communication  & 0.2k  & 154 KB       \\
CovidTracker      & Tool          & 1.6k    & 2.6 MB       & LayoutAnimation      & Tool           & 0.2k  & 102 KB       \\
CovidTrackerKen   & Tool          & 0.6k    & 12.0 MB      & Logger               & Productivity   & 1.1k  & 157 KB       \\
ESFileExplorer    & Tool          & 1096.0k & 54.2 MB      & Meditime             & Medicine       & 1.5k  & 276 KB       \\
EyeWitness        & Communication & 3.3k    & 3 MB         & MobileSports         & Sports         & 3.2k  & 11.8 MB      \\
FastLib           & Productivity  & 16.4k   & 49.6 MB      & MultiChate           & Communication  & 1.3k  & 192 KB       \\
iWeather          & Weather       & 4.1k    & 12.6 MB      & PrzepisyApp          & Tool           & 1.2k  & 2.6 MB       \\
Kassenschnitt     & Finance       & 1.2k    & 250 KB       & Purples              & Tool           & 2.5k  & 12.4 MB      \\
NeverTooManyBooks & Reading       & 110.6k  & 10.1 MB      & QuizApp              & Education      & 0.9k  & 137 KB       \\
PopularMovies     & Entertainment & 2.5k    & 4.6 MB       & roboTV               & Entertainment  & 38.7k & 11.1 MB      \\
ProgressNote      & Productivity  & 4.1k    & 17.6 MB      & SimpleTorch          & Tool           & 0.2k  & 1.7 MB       \\
SakuraAnime       & Entertainment & 9.8k    & 9.1 MB       & SuperHeroInteraction & Entertainment  & 3.3k  & 3.0 MB       \\
TEFAPTracker      & Tool          & 1.1k    & 506 KB       & TextToSpeech         & Tool           & 1.9k  & 250 KB       \\
Unter             & Sports        & 4.7k    & 8.8 MB       & ToDay                & Productivity   & 8.4k  & 26.1 MB      \\
VocableTrainer    & Education     & 15.3k   & 2.5 MB       & ToDoApp              & Productivity   & 7.4k  & 257 KB       \\
AdvancedHacker    & Tool          & 1.2k    & 164 KB       & ToDoNotes            & Productivity   & 7.9k  & 19 MB        \\
Agility           & Tool          & 148.6k  & 26.2 MB      & UILog                & Tool           & 0.4k  & 149 KB       \\
AMVP              & Tool          & 0.3k    & 47 KB        & UniPool              & Transportation & 4.6k  & 1.0 MB       \\
BMI               & Tool          & 0.2k    & 796 KB       & WearLocker           & Tool           & 0.5k  & 155 KB       \\
BuildItBigger     & Productivity  & 3k      & 50.8 MB      & WordNotebook         & Productivity   & 4.5k  & 12.0 MB \\ \bottomrule     
\end{tabular}}
\end{table}

From the apps satisfying the above two criteria, we choose 50 apps with the highest popularity (project repository star numbers). We recruit three graduate students to develop \vvv{10} scripts for each AUT (in total \vvv{500} scripts). The numbers of operations contained in the test scripts ranges from 1 to 16. The length of the test scripts corresponds to the actual situation of test scripts in real-world industry. The \vvv{500} test scripts have an average of 4.65 operations. The standard deviation is 2.76. \vvv{The students are not involved in the user study.} The students are with at least three years of experience in mobile app testing \vvv{in the real industry environment}. They are familiar with Appium and Appium scripts for mobile app testing. During their script development, it is required that each script should focus on one single and clear target functionality or test goal, which can unambiguously express the user expectations of app behaviors for specific testing operations, \ie test intention. \vvv{Moreover, it is required that different test scripts should cover different functionalities of the AUT, so that the generalizability of the \toolname can be guaranteed. According to our manual check, our test scripts in the experiment cover the vast majority of the main functionalities of the apps.} The students are provided with real-world test scripts from the industry as examples and guidance. Besides, they are provided with templates that they only need to fill in the test operation part of the test scripts, and the initialization part of the \texttt{AppiumDriver} is provided. \vvv{Combined with the actual development and testing experience of the students from real industry environment, we can try our best to reduce the potential bias of the experiment scripts with the scripts from actual testers in commercial settings.}

\vvv{The three students are further required to label the ground-truth intentions of the test scripts and all the operations. Each test script and each test operation are labeled with an unambiguous test intention. The three students are required to label the ground-truth intentions independently. They refer to the business logic in official documents, the code annotations, and the GUI information of the apps to do the labeling work. For the three results from three students on the same test script, we do a Fleiss’ kappa test, and if the value is larger or equals to 0.9, which indicates ``almost perfect agreement'', they will discuss to reach an agreement. For those test scripts (which is only 12 in our experiment) whose Fleiss’ kappa test value is less than 0.9, we invite two third-party experts (one from academia and two from industry) who have over 10 years of mobile app development and testing to decide the final ground-truth labels, and we believe their labeling is convincing.}

\subsubsection{\vvv{Baseline Approach}}

\vvv{
In our evaluation, we use two groups of baselines. The first group is representative and state-of-the-art source code comment generation approaches, including \cdsq \cite{alon2018code2seq}, \dome \cite{mu2023developer}, and \llm\footnote{An large language model-based approach with no official tool name} \cite{geng2024large}, and the second group is \gui and \code, which are parts of \toolname, as an ablation experiment.

\cdsq \cite{alon2018code2seq}: The \cdsq approach is one of the state-of-the-art and most representative approaches for source code comment generation, which has been proven effective for Java code summarization. It is also representative of a series of state-of-the-art learning-based code comment generation approaches. We directly input the test scripts into \cdsq to infer the test script intentions to evaluate its capability on test script intention generation. 

\dome \cite{mu2023developer}: The \dome is another state-of-the-art and representative source code comment generation approach. It utilizes Intent-guided Selective Attention to explicitly select intent-relevant information from the source code, and produces various comments reflecting different intents.

\llm \cite{geng2024large}: The \llm is based on the capabilities of large language models (LLMs). It adopts the in-context learning paradigm and giving adequate prompts to the LLMs. It also uses customized strategies for constructing the prompts and post-processing strategies for reranking the results.

\gui and \code are different parts of \toolname, which can extract textual or image information from the GUI layout files and GUI images, and can map operations to the response methods and then generate natural language descriptions for the test script with a sequence-to-sequence model. We directly apply \gui and \code to evaluate its capability on test script intention generation for the widgets located both by \id and \xpath, respectively. These two baselines are considered as the ablation experiments of \toolname, and can indicate whether these two parts are both important to \toolname.
}

\subsubsection{\textbf{Evaluation Metric}}

\toolname can be considered a sequence-to-sequence model, which transfers a series of code sequences to a series of natural language sequences. Therefore, we adopt the widely used evaluation metrics in sequence-to-sequence tasks, BLEU \cite{papineni2002bleu}, CIDEr \cite{vedantam2015cider}, METEOR \cite{banerjee2005meteor} and ROUGE \cite{lin2003automatic}.

\textbf{BLEU} (\underline{B}i\underline{L}ingual \underline{E}valuation \underline{U}nderstudy) is calculated as the product of $N$-gram precision and brevity penalty, where the $N$ can take any positive integer. \vvv{It measures the similarity between one sentence to a set of reference sentences using constituent n-grams precision scores.} For the reason that most test script intentions are short, we take $N$ equaling 1, 2, 3, and 4, and denote them as BLEU@1, BLEU@2, BLEU@3, and BLEU@4. \vvv{It can be calculated as:} 

\begin{align}
\small
\vvv{\text{BLEU} = \left( \begin{cases} 1 & \text{if } c > r \\ \exp(1 - \frac{r}{c}) & \text{if } c \leq r \end{cases} \right) \cdot \exp \left( \frac{1}{n} \sum_{n=1}^{n} \log \left( \frac{\sum_{C \in \{Candidates\}} \sum_{n\text{-gram} \in C} \text{Count}_{\text{clip}}(n\text{-gram})}{\sum_{C \in \{Candidates\}} \sum_{n\text{-gram} \in C} \text{Count}(n\text{-gram})} \right) \right)}
\end{align}

\vvv{where $c$ is the length of the candidate translation, $r$ is the length of the reference translation that is closest in length to the candidate, $N$ represents the maximum length of n-grams considered in the calculation, $\text{Count}(n\text{-gram})$ is the number of times an n-gram appears in the candidate translation, and $\text{Count}_{\text{clip}}(n\text{-gram})$ is the maximum number of times an n-gram appears in any reference translation.}

\textbf{CIDEr} (\underline{C}onsensus-Based \underline{I}mage \underline{D}escription \underline{E}valuation) \vvv{is a metric designed to assess the quality of image captions by comparing them to multiple human-generated references. It evaluates captions based on consensus, using term frequency-inverse document frequency (TF-IDF) weighting to account for the importance of words, and measures how closely a generated caption matches the reference captions in terms of content and relevance. CIDEr emphasizes capturing the consensus of human judgments, making it effective for evaluating the semantic correctness and informativeness of image descriptions.} \vvv{It can be calculated as:} 

\begin{equation}
\vvv{\text{CIDEr}_n(c, S) = \frac{1}{|S|} \sum_{s \in S} \frac{ \sum_{g \in G} \text{TF-IDF}_g(c) \cdot \text{TF-IDF}_g(s) }{ \sqrt{ \sum_{g \in G} (\text{TF-IDF}_g(c))^2 } \cdot \sqrt{ \sum_{g \in G} (\text{TF-IDF}_g(s))^2 }}}
\end{equation}

\vvv{where $c$ is the candidate sentence, $S$ is the set of reference sentences, $G$ is the set of n-grams, and $\text{TF-IDF}_g(x)$ is the TF-IDF score of n-gram $g$ in text $x$.}

\textbf{METEOR} (\underline{M}etric for \underline{E}valuation of \underline{T}ranslation with \underline{E}xplicit \underline{OR}dering) is widely used for the sequence-to-sequence model evaluation. Compared with BLEU, METEOR takes the synonyms and recall ratio into consideration. \vvv{It evaluates the generated summary by aligning it to the reference summary and calculating the similarity scores based on the unigram matching.} \vvv{It can be calculated as:} 

\begin{equation}
\vvv{P = \frac{N_m}{N_{tc}}, \quad R = \frac{N_m}{N_{tr}}, \quad \text{METEOR} = \frac{10 \cdot P \cdot R}{P + 9 \cdot R} \cdot \left( 1-0.5 \cdot \frac{N_c}{N_m} \right)}
\end{equation}

\vvv{where $N_m$ refers to number of matched unigrams, $N_{tc}$ refers to total number of unigrams in candidate sentences, $N_{tr}$ refers to total number of unigrams in reference sentences, and $N_c$ refers to number of contiguous matched sequences.}

\textbf{ROUGE} (\underline{R}ecall-\underline{O}riented \underline{U}nderstudy for \underline{G}isting \underline{E}valuation) is based on the recall ratio. \vvv{It computes the count of several overlapping units such as n-grams, word pairs, and sequences.} We use a variant of ROUGE, ROUGE-L, which calculates the similarity between candidate sentences generated by \toolname and reference sentences based on the longest common subsequence (LCS). \vvv{It can be calculated as:} 

\begin{equation}
\vvv{\text{ROUGE-L} = \frac{(1 + \beta^2) \cdot LCS(C, R)}{|R| + \beta^2 \cdot |C|}}
\end{equation}

\vvv{where $LCS(C, R)$ is the length of the LCS between candidate sequence $C$ and reference sequence $R$, $|C|$ and $|R|$ are the total number of words in the candidate sequence and reference sequence, and $\beta$ is a parameter that determines the weight of recall relative to precision, usually $\beta$=1 used.}

The range of all the adopted metrics is $[0, 1]$. The higher the score is, the closer the \toolname generated intentions are to the ground-truth intentions. 

\subsection{RQ1: Intention Generation}

To evaluate the effectiveness of \toolname on generating test intentions, we compare \toolname with three baselines.

\begin{table}[!h]
\caption{\vvv{Experimental Results and Comparison}}
\label{tab:result}
\centering
\scalebox{0.9}{
\begin{tabular}{cccccccc}
\toprule
& BLEU@1 & BLEU@2 & BLEU@3 & BLEU@4 & CIDEr & METEOR & ROUGE \\ \midrule
code2seq             & 0.05    & 0.01    & 0.00    & 0.00    & 0.06  & 0.06   & 0.08     \\
DOME                 & 0.19    & 0.04    & 0.00    & 0.00    & 0.12  & 0.20   & 0.20     \\
LLM-basis            & 0.15    & 0.03    & 0.00    & 0.00    & 0.16  & 0.25   & 0.20     \\ \midrule
GUIIntention         & 0.43    & 0.06    & 0.02    & 0.01    & 0.12  & 0.21   & 0.18     \\
CodeIntention        & 0.10    & 0.01    & 0.00    & 0.00    & 0.07  & 0.06   & 0.06     \\ \midrule
TestIntention (step) & 0.59    & 0.38    & 0.26    & 0.15    & 0.24  & 0.33   & 0.19     \\
TestIntention        & 0.45    & 0.17    & 0.05    & 0.02    & 0.18  & 0.29   & 0.25  \\ \bottomrule   
\end{tabular}}
\end{table}

\tabref{tab:result} shows the experimental results. The average BLEU score ($N=1, 2, 3, 4$) of \toolname on the script granularity is \vvv{0.17}. The scores decrease with the $N$ increases. The reason is that for the operation sequence, a few words can summarize the test intention, and the ground-truth sentences are short. When $N$ increases, it is more likely that the generated intentions differ from the ground-truth intentions. For another three metrics, the CIDEr score is \vvv{0.18}, the METEOR score is \vvv{0.29}, and the ROUGE score is \vvv{0.25}.

By comparison, \toolname behaves much better than \vvv{the baseline approaches, including the LLM-based one}. One reason is that in test scripts, there contains little or no business logic of the app, and the scripts only contain the operation sequences where app widgets are located by \ids and \xpaths. Therefore, the baseline approaches cannot extract effective information from the test scripts. \toolname performs better via mapping each operation to the response method or the GUI image information, which helps take the business logic information into consideration and effectively helps the intention generation.

Also, \toolname behaves better than the \gui and the \code on all of the evaluating metrics. That is to say, the information from only GUI layout files or from only response methods are not adequate for the test intention generation. The reason is that during the development of test scripts, developers will choose different selectors to identify the corresponding widgets, while neither \gui nor \code can singly obtain enough information that can be used to infer test script intentions. In order to better obtain the test script intentions, it is necessary to combine the GUI information and the response method information. Further, \gui outperforms \code. One reason is that most information can be obtained via \gui than \code (more detailed analysis in \secref{sec:rq2}). Moreover, the test intentions of specific operations are explicitly or implicitly presented to app end users, and information obtained from GUI with \gui can assist in inferring the test intentions more accurately than \code. \vvv{According to the ablation experiment results, we confirm that both parts of the \toolname are important to the \toolname as a whole.}

Further, we analyze the generated intentions at the step granularity to show the capability of \toolname in generating of single operations. \vvv{500} test scripts contain \vvv{2323} steps, and the average BLEU score ($n=1, 2, 3, 4$) is \vvv{0.34}, \vvv{99.76\%} higher than the integrated results. The CIDEr score is \vvv{0.24}, \vvv{29.15\%} higher than the integrated results. The METEOR score is \vvv{0.33}, \vvv{13.77\%} higher than the integrated results. The ROUGE score is \vvv{0.19}. ROUGE of the step granularity is lower than integrated results because that the calculation of ROUGE involves the LCS calculation, while the step granularity results may limit the common subsequence length.

\begin{lstlisting}[
	language = Java, 
	caption  = Generation Example of \toolname in RQ1, 
	label    = code:result
]
public class ProgressNote_1 {
  // ground-truth: add a note
  // TestIntention: add new note, title, and content
  public void createNote(AppiumDriver driver) {
    //initialization work
    ...
    // ground-truth: add new note
    // TestIntention: add note
    driver.findElementById("cn.zerokirby.note:id/floatButton").click();
    Thread.sleep(1000);
    // ground-truth: add title
    // TestIntention: add title
    driver.findElementById("cn.zerokirby.note:id/noteTitle").sendKeys("One Day");
    Thread.sleep(1000);
    // ground-truth: add content
    // TestIntention: add content
    driver.findElementById("cn.zerokirby.note:id/mainText").sendKeys("It's a really fine day!");
    // ground-truth: add extra content
    // TestIntention: add
    driver.findElementById("/hierarchy/FrameLayout/LinearLayout/.../ViewGroup/TextField")
      .sendKeys("Another fine day!");
    Thread.sleep(1000);
    // ground-truth: save
    // TestIntention: 
    driver.findElementById("cn.zerokirby.note:id/save").click();
    Thread.sleep(1000);
  }

  public static void main(String[] args) {
    // other configurations
    ...
  }
}
\end{lstlisting}

\coderef{code:result} shows a real GUI test script from our evaluation. It can be found that the intention of the script, and the intentions of the first, second, third, and fifth steps are successfully generated, but the generation of the intention of the fourth step failed.

\subsection{RQ2: Operation Mapping}
\label{sec:rq2}

\begin{figure}[!h]
\centering
\includegraphics[width=0.7\linewidth]{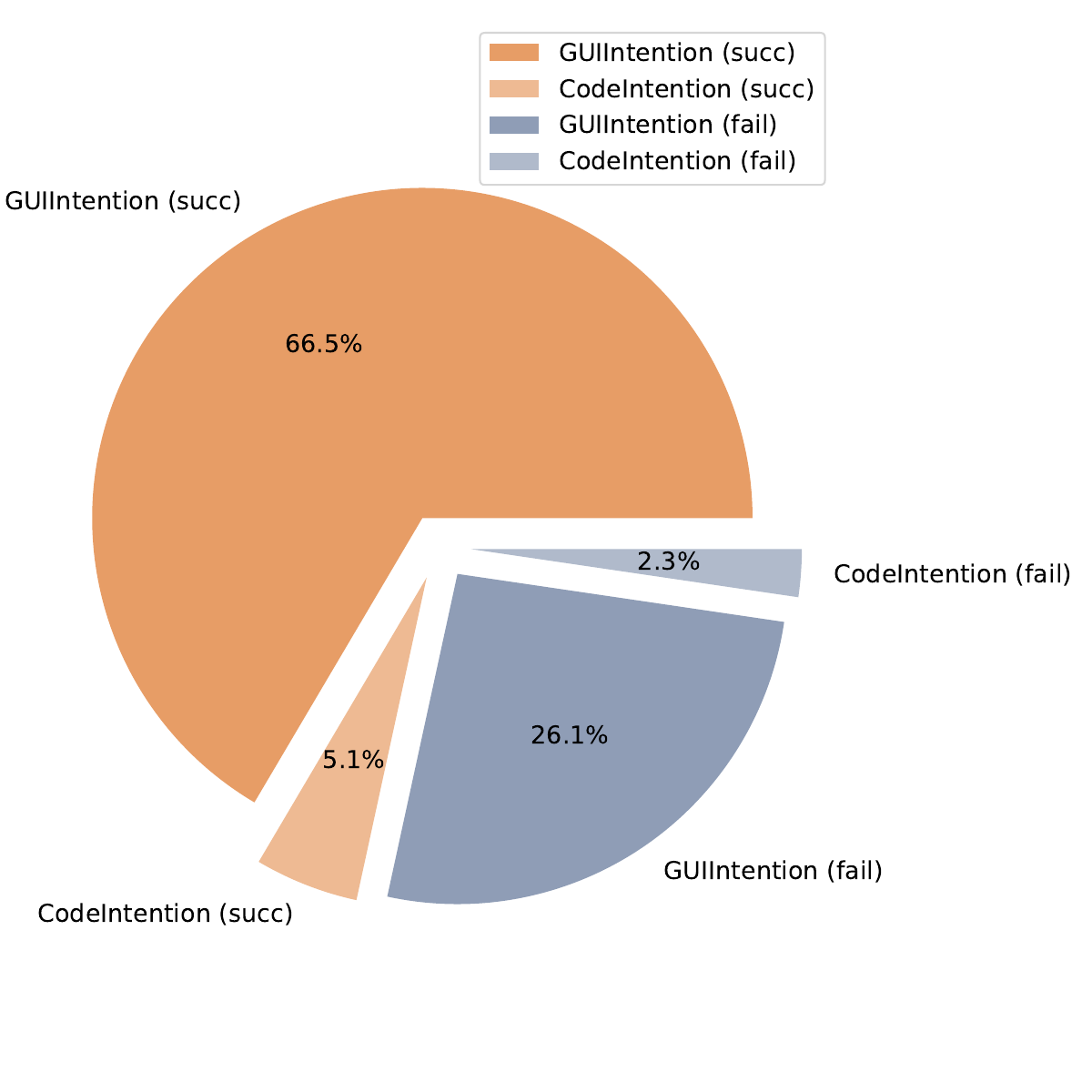}
\vspace{-1cm}
\caption{\vvv{Result of Operation Mapping Effectiveness}}
\label{fig:RQ2}
\end{figure}

Operation mapping is one of the most important parts and most noteworthy contributions of \toolname. This research question is set to verify the effectiveness of \toolname to map operation to corresponding information, including response method and GUI information. This research question is studied on the step granularity because the operation mapping is on the step granularity. \vvv{1663} out of \vvv{2323} steps are successfully mapped \vvv{(71.6\%)} to the expected information (\figref{fig:RQ2}).

Among the successful mapping operations \vvv{(1663)}, intentions of about \vvv{93\%} operations \vvv{(1545)} are generated by \gui. The main reason is that \ids are not always available during the test script development. The usage of \ids heavily depends on the app developers' habits on whether they label the widgets with \ids. Therefore, an unstandardized app development process leads to the \id missing. Second, the Android mechanism allows developers to use repetitive \ids, making test developers unable to identify the target widgets with a unique \id. Such a practice prevents \toolname from mapping the operation to the appropriate response methods. For manual effort, it is even for human testers to effectively identify the response methods via the limited \id information provided in the app source code. On such a condition, \gui is especially critical, which is a prominent contribution of this paper. \vvv{Besides, for all the operation mapping, including successful ones and failed ones, 21.4\% are located by \ids (497), among which 142 are failure, and 78.6\% are located by \xpaths (1826), among which 518 are failure; 92.6\% are mapped by the \gui (2151), 7.3\% and are mapped by the \code (172).}

We also conduct an investigation on the failing results to see why some operations cannot be mapped by \id or by \xpath. Among the failing mapping operations, according to our in-depth observation, most operations that fail to be mapped are transition operations. In other words, such operations do not have concrete test intentions, and they just play the role of transiting from the former operation to the next operation. For example, \texttt{scroll()} function just means to scroll the app activity to link the former and latter operations to push forward the testing process, and it does no contains the target functionalities or test goals, \ie the test intention concept defined in this paper. Such transition operations do not contain specific test intentions, so it is hard to infer the corresponding test intentions for transition operations even for testing experts. \vvv{As we described in our motivation part, there are only selector type, selector, operation, and content in the test scripts, it is hardly possible to identify which steps are transition operations, and such transition operations are identified during our manual review to the evaluation results, and it is also difficult to automatically generate ``intentions'' for such transition operations. Besides, such transition operations do not actually express any ``intentions'' targeting on the app functionalities. However, such transition operations naturally exist in the GUI test scripts, we believe it may cause a bias if we directly remove such transition operations during our evaluation.}

\subsection{RQ3: A User Study on User Experience}

\vvv{To better illustrate the usefulness of \toolname, we further conduct a user study. We randomly choose 75 test scripts out of our test script dataset for 15 other recruited participants (not involved in the script labeling). The participants are familiar with mobile app testing and Appium scripts, and the \toolname is transparent to them. 6 of the participants are male and 9 are female. 10 of the participants are second-year master students majoring in software engineering and 5 are final-year undergraduate students majoring in software engineering. All of the participants have at least two years of experience in mobile app testing. A significance test on the participants’ capability in mobile app test script development are conducted, and the significance value is smaller than 0.05, showing there is no significant difference in capabilities among the three participants. Based on the 75 test scripts, we construct two test script sets. $Set A$ is the initial test scripts without test intentions. $Set B$ is the test scripts attached with \toolname generated test intentions. The test scripts are labeled from $A1$ to $A75$, and $B1$ to $B75$. The 150 test script items are randomly allocated to the 15 participants. Each participant is allocated with 10 test scripts, with 5 from $Set A$ and 5 from $Set B$. We ensure that the 10 scripts are different and no participants will receive both labelled and unlabeled versions of one same script. The test script allocation is shown in \tabref{tab:userstudy}.

\begin{table}[!h]
\caption{\vvv{Test Script Allocation in the User Study}}
\label{tab:userstudy}
\centering
\scalebox{0.9}{
\begin{tabular}{c|ccccccccccccccc}
\toprule
   & P1  & P2  & P3  & P4  & P5  & P6  & P7  & P8  & P9  & P10 & P11 & P12 & P13 & P14 & P15 \\ \midrule
1  & A10 & A8  & A23 & A12 & A9  & A15 & A54 & A31 & A1  & A6  & A3  & A2  & A11 & A4  & A5  \\
2  & A18 & A16 & A24 & A19 & A30 & A29 & A63 & A32 & A20 & A22 & A7  & A38 & A26 & A43 & A21 \\
3  & A46 & A17 & A27 & A52 & A35 & A37 & A65 & A44 & A25 & A34 & A13 & A53 & A28 & A50 & A42 \\
4  & A48 & A40 & A67 & A59 & A36 & A41 & A70 & A51 & A47 & A45 & A14 & A61 & A33 & A55 & A49 \\
5  & A56 & A74 & A69 & A64 & A71 & A57 & A73 & A62 & A58 & A60 & A39 & A72 & A68 & A66 & A75 \\ \midrule
6  & B14 & B18 & B5  & B9  & B33 & B6  & B2  & B1  & B4  & B3  & B40 & B7  & B8  & B21 & B26 \\
7  & B16 & B46 & B11 & B23 & B60 & B22 & B13 & B20 & B19 & B10 & B47 & B17 & B12 & B27 & B28 \\
8  & B32 & B55 & B30 & B24 & B65 & B35 & B38 & B25 & B31 & B15 & B50 & B29 & B62 & B37 & B45 \\
9  & B42 & B57 & B44 & B36 & B69 & B39 & B51 & B52 & B34 & B59 & B61 & B43 & B64 & B41 & B53 \\
10 & B72 & B56 & B48 & B67 & B73 & B74 & B58 & B54 & B63 & B75 & B71 & B68 & B70 & B49 & B66 \\ \bottomrule
\end{tabular}}
\end{table}

In order to determine the usefulness of \toolname, we set a task for the participants to complete, and we record the time they use. The specific task is that the participants are required to manually operate on the apps to go through all the target app activities according to the test operations described in the test scripts. This task can reflect whether the participants understand the test scripts and can find the corresponding functions in the app. This is also a prerequisite of script update and repair in real-world test script development and maintenance. The time calculation starts from the opening of the test script file, and ends with the app transiting to the final correct target app activity. One thing to notice is that we require all the activities should be strictly consistent to the operations recorded in the test scripts, and if not, the time calculation will not end. In order to help the participants better understand the task, we provide three pairs of test scripts (labelled and unlabeled versions) and the corresponding apps in advance (no overlap with the scripts used in the user study). They have adequate time to get familiar with the scripts and apps, and the specific task.}

\begin{figure}[!h]
\centering
\includegraphics[width=\linewidth]{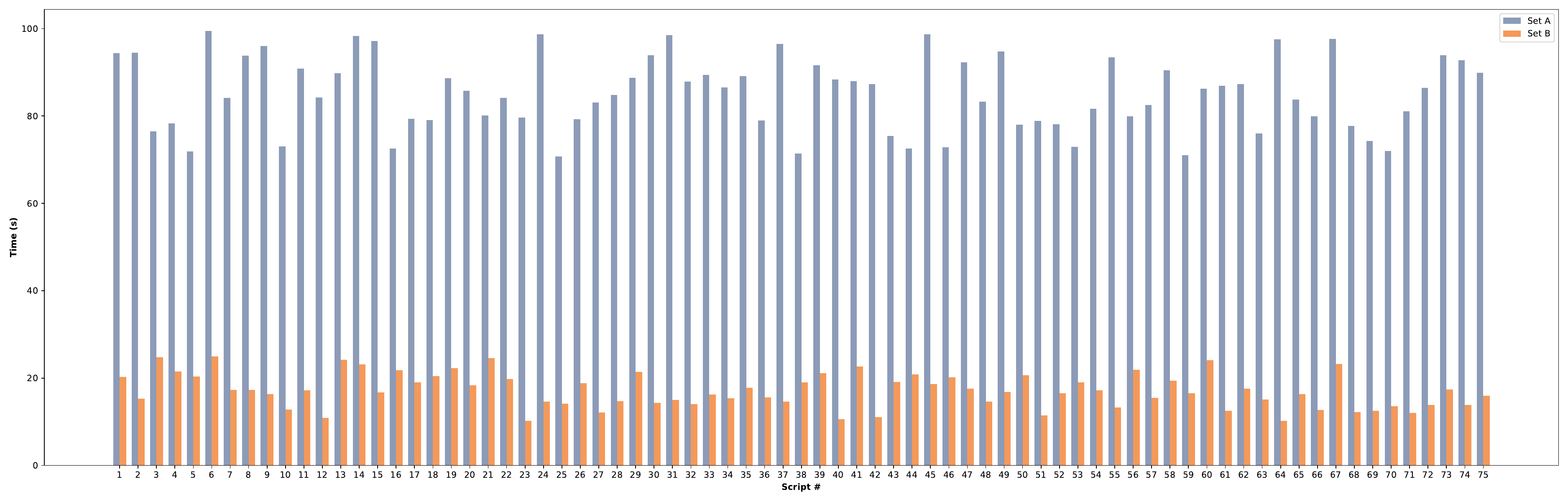}
\caption{\vvv{Result of User Study}}
\label{fig:RQ3}
\end{figure}

The results can be seen from \figref{fig:RQ3}. The average \vvv{task completion} time for $Set A$ is \vvv{85.13} seconds, and the average time for $Set B$ is \vvv{17.13} seconds. With the assistance of \toolname, when faced with unfamiliar mobile app GUI test scrips, testers can save \vvv{79.88\%} of the time, and testers need to spend \vvv{4.97} times of time overhead without \toolname. The data show that \toolname can effectively improve the test script development and understanding efficiency. On different test scripts, the improvement ranges \vvv{from 10.47\% to 32.41\%}, and the spent time reaches \vvv{3.09 to 9.55} times of time overhead, depending on the test script complexity and the business logic of the app under test. The outstanding performance of \toolname shows its capability to assist developers in understanding test scripts. \toolname helps save time by reducing the time on manually mapping the test operations to app source code or GUI information. The usability of \toolname is confirmed by the user study.

\subsection{\vvv{Evaluation on Used Models}}

\vvv{In this section, we present the evaluation on the used models in our approach \toolname.}

\subsubsection{\vvv{Widget Image Understanding}} \vvv{The widget image understanding model is evaluated on a dataset containing 5000 different widget images. Such widget images are collected from the widely used Rico dataset\footnote{http://www.interactionmining.org/rico.html}. After obtaining the widget images, we recruit three graduate students to label the images with their intentions, who are also recruited for the \toolname intention generation ground-truth labeling. The demographic background of the students is presented in \secref{sec:dp}. Following a similar process, the three students are required to label the ground-truth intentions independently. For the three results from three students on the same widget image, we do a Fleiss’ kappa test, and if the value is larger or equals to 0.9, which indicates ``almost perfect agreement'', the three results will be kept. For those widget images (which is only 121 in this evaluation) whose Fleiss’ kappa test value is less than 0.9, we invite three third-party experts (one from academia and two from industry) who have over 10 years of mobile app development and testing to decide the final ground-truth labels, and we believe their labeling is convincing. Due to the fact that the intentions of widget images are all short phrases with one or two words, we use the exact match. synonyms are also considered correct by manual check. As a result, the intentions of 4873 widget images are correctly generated, which takes a percentage of 97.46\%, while the original model only correctly generates the intentions of 3917 widget images (78.34\%). The results can well indicate the effectiveness of the widget image understanding model.
}

\subsubsection{\vvv{Code Intention Generation}} \vvv{The code intention generation model is evaluated on a dataset containing 3000 source code fragments. Following the similar process described in \secref{sec:dp} (we omit the process description because these two processes are the same), we obtain the ground-truth labels describing the code intentions for each source code fragment. We also use the same metrics BLEU \cite{papineni2002bleu}, CIDEr \cite{vedantam2015cider}, METEOR \cite{banerjee2005meteor} and ROUGE \cite{lin2003automatic}, with comparison to the same baselines \cdsq \cite{alon2018code2seq}, \dome \cite{mu2023developer}, and \llm \cite{geng2024large}. The results are shown in \tabref{tab:model}.
}

\begin{table}[!h]
\caption{\vvv{Code Intention Generation Model Effectiveness}}
\label{tab:model}
\centering
\scalebox{0.9}{
\begin{tabular}{cccccccc}
\toprule
& BLEU@1 & BLEU@2 & BLEU@3 & BLEU@4 & CIDEr & METEOR & ROUGE \\ \midrule
code2seq             & 0.07 & 0.03 & 0.01 & 0.00 & 0.11 & 0.08 & 0.12 \\
DOME                 & 0.21 & 0.03 & 0.01 & 0.00 & 0.15 & 0.23 & 0.31 \\
LLM-basis            & 0.24 & 0.17 & 0.09 & 0.01 & 0.21 & 0.22 & 0.33 \\ 
TestIntention model  & 0.53 & 0.29 & 0.19 & 0.11 & 0.28 & 0.38 & 0.41 \\ \bottomrule   
\end{tabular}}
\end{table}

\vvv{From \tabref{tab:model}, we can see that after fine-tuning the \cdsq model with mobile apps, the code intention generation model is more effective. Compared with different baselines, the improvement of the code intention generation model of \toolname over BLEU@1, BLEU@2, BLEU@3, BLEU@4, CIDEr, METEOR, ROUGE reaches 24.24\% - 6683.60\%. 
}

\subsection{Threats to Validity}

Some factors may pose threats to the validity of the experiment, while some settings can help relieve the influence.

\textbf{The apps we use in the experiment might be limited.} We use \vvv{50} apps in this experiment, which seems inadequate. However, the apps cover different categories, including tools, productivity, finance, entertainment, \etc Besides, during the development of the test scripts, we require the students to cover as many scenarios (functionalities) as possible to ensure the generalizability of \toolname. Moreover, the only restrictions have been claimed in \secref{sec:dp}. Therefore, we hold that the apps are representative and can well prove the generalizability and scalability of \toolname.

\textbf{The test scenarios we set in the experiment are limited.} For each app, we require the graduate students to set \vvv{10} different test scripts, which represent five different test scenarios. The test scripts do not cover all the source code and do not fully explore the functionality space. However, during the development of the test scripts, the students are required to refer to the apps' documents and follow the dominating business logic paths to set the test scenarios, which are representative. \vvv{Moreover, different test scripts are required to cover different functionalities.}

\textbf{The manual efforts involved in this experiment are graduate students.} During the experiment and the preliminary survey (\secref{sec:siu}), we recruit senior software engineering major students to design the test scenarios and develop the test scripts. This might be a potential threat. However, Salman \etal propose that senior students are sufficient developer proxies in well-controlled experiments \cite{salman2015students}. Besides, we provide examples from real-world app test scripts in the industry and the script template, with irreverent parts being completed. This can maximize the reliability of the scripts in the experiment as agents of real-world scripts and minimize the effect of the personal coding style of the recruited students. Therefore, such students can effectively develop diverse test scripts that cover many different scenarios on different app functions.

\vvv{\textbf{There may exist multiple instances of a single component in real-world practice of Android app development.} However, we tried to eliminate such a threat. First, we use a ranking algorithm for different templates, which can help reduce the potential threat. Second, according to our manual review on the actual apps, we find that although some apps exist multiple instances of a single component, which may be caused by dynamic widget generation during runtime, such widgets are always with the same intentions, which will not affect our target of TestIntention. Third, we manually checked all the evaluation results, and we find that such a situation is quite rare in actual practice. Therefore, we believe such a situation will not pose a threat to our approach.}

\vvv{\section{Discussion}}

\vvv{
In our work, we introduce \toolname to generate GUI test script intentions in the form of natural language. It is expected that app developers can better understand the intentions with unfamiliar GUI test scripts. Alternatively, specific languages or syntaxes are supposed to be able to specify tests in a more structured way, like the Gherkin syntax according to Behavior-Driven Development (BDD)\footnote{https://cucumber.io/docs/gherkin/}. However, compared natural language descriptions, such specific syntaxes still have some limitations. First, such syntaxes are limited when describing complex business logic or details with the keywords and structures, while natural language can more flexibly describe complex issues. Second, app developers still need to additionally learn the syntaxes, which may lead extra human efforts and time overhead. App developers are required to learn the basic syntaxes and the specific usages in specific app testing scenarios. Third, the structures of the specific languages or syntaxes are comparatively fixed, which may restrict the flexibility. Therefore, considering the above factors, we choose to generate GUI test script intentions in the form of natural language, making \toolname more practical.

Furthermore, as we described above, \toolname generates GUI test script intentions, which refers to the user expectations of app behaviors for specific operations. Such intentions may implicitly contain some of the test result expectations of the test scripts. The target of this work is to better facilitate app developers during the app GUI test script development and maintenance, so \toolname focuses on the targets of GUI test scripts and corresponding steps. However, we do not formally contain the test oracles in our work. We believe it is an interesting and meaning future research direction of app GUI script understanding, and will be a strong improvement of our work. As the first work focusing on the app GUI script understanding as far as we can know, we believe that our work is a significant step towards the app test script development and maintenance, and the involvement of test oracle understanding will be a meaningful future direction.}

\section{Related Work}

\subsection{Source Code Summarization}

Code comment generation, also known as code summarization, has been a long-lasting and important topic in the program analysis field. Such approaches can be seen as inferring the intentions of the source code. Main-stream code summarization approaches include three major categories, manual template, information retrieval (IR), and learning-based approaches.

As a representative work of the manual template approach, Sridhara \etal propose a method for automatically generating descriptive comments for Java methods with the analysis on the AST and call flow graph (CFG) \cite{sridhara2010towards}. 
Another work of Sridhara \etal focuses on the high-level actions \cite{sridhara2011automatically}. The approach analyzes the statement blocks and generates comments for each high-level action. 
As a representative work of the manual template approach, Sridhara \etal propose a method that focuses on the high-level actions \cite{sridhara2011automatically}. The approach analyzes the statement blocks and generates comments for each high-level action. 
Rui \etal conclude 26 different ``nano-patterns'' \cite{rai2017method}, which are used as templates to generate code comments. 
Moreno \etal introduce a class-level heuristic code summarization approach \cite{moreno2013automatic}, focusing on the class content and responsibilities. 
Malhotra \etal also generate code comments from class level \cite{malhotra2018class}. They analyze the dependency relationship and analyze the change proneness. 
Nazar \etal recruit students to label the collected projects based on 21 different features and construct a classifier to generate code comments \cite{nazar2016source}.

IR models are also widely used in code comment generation models. 
Haiduc \etal first extract the words from code files and link them with the most related words in the corpus and generate code comments \cite{haiduc2010use}.
Haiduc \etal also propose a method to generate code comments based on code lexical information and structure information \cite{haiduc2010supporting}.
Haiduc \etal first extract the words from code files and link them with the most related words in the corpus and generate code comments \cite{haiduc2010use, haiduc2010supporting}.
Interestingly, Rodeghero \etal identify the statements and keywords developers focus on based on eye-tracking technologies \cite{rodeghero2015eye, rodeghero2014improving}. Then, based on the information acquired, they can generate higher-quality code comments. 
Some techniques focus on software repository mining.
Wong \etal utilize crawling technology to obtain a large amount of code snippets and descriptions \cite{wong2013autocomment}, construct the ``code-description'' mappings as the corpus, and use code similarity to infer the code comments.
Alon \etal propose \cdsq and Code2Vec \cite{alon2018code2seq, alon2019code2vec}, considering the semantic similarity among methods, to generate code comments.

Taking advantage of the development of DL technology, code comment generation also has a breakthrough.
Iyer \etal propose CODE-NN \cite{iyer2016summarizing}, which combines LSTM and Attention mechanism to encode and decode the code snippets.
Zheng introduces a novel attention mechanism, code attention \etal \cite{zheng2017code}, to use domain features (\ie symbol, identifier) to understand the code structure.
Hu \etal transfer code snippets into AST to analyze the structure and semantic information \cite{hu2018deep}. 
Leclair \etal propose ast-attendgru \cite{leclair2019neural}, which combines code text and AST representation as neural network input to generate code comments.
Huang \etal \cite{huang2023bcgen} propose a bytecode comment generation approach using a neural language model to understand the meaning of the bytecode and further help developers in more software activities.
Gao \etal \cite{gao2023code} propose a novel approach named SG-Trans to incorporate code structural properties into Transformer, and then to generate accurate code summaries.
Zhou \etal \cite{zhou2023towards} propose a new framework, MLCS, for code summarization based on meta-learning and code retrieval. MLCS forms the summarization of each code as a few-shot learning task, where similar examples are used as training data and the testing example is itself.

Such work would have a deep investigation on source code understanding or comment generation \cite{fang2022prhan, zhou2022summarizing}, and the adopted algorithms are valuable to the \toolname construction, since after mapping test operations to source code, it is necessary to understand the business logic of source code. Some work \cite{daka2017generating, li2018aiding, li2016automatically, roy2020deeptc, panichella2016impact, zhang2016towards} also tries to understand white-box unit tests with NLP technologies. However, test scripts are still not well investigated to understand the test script intentions. \vvv{Compared with source code summarization or source code comment generation tasks, GUI test script intention generation is more complex. However, as far as we can know, there is no related studies focused on the code understanding or code summarization for test cases or test scripts. As a result, we build our work on the basis of source code comment generation approaches and compare our work with such approaches. The main difference between test case understanding and source code understanding is that the source code contains intentions while test cases are not explicitly or directly linked with the business logic, and thus mapping from tests to business logic is an important prerequisite.}

\subsection{GUI Image Analysis}

With the development of computer vision and deep learning, GUI image analysis has advanced. 

Some work analyzes the GUI images and uses the information to reconstruct the editable GUI files.
Chen \etal present a neural network machine translator that combines recent advances in computer vision and machine translation and translates GUI images into GUI skeletons \cite{chen2018ui}.
Pix2Code \cite{beltramelli2018pix2code} is a method presented by Beltramelli. It applies an end-to-end image captioning model to predict a description of the GUI layout.
Chen \etal introduce a novel approach that automates cross-platform GUI code generation with the detection and extraction of GUI widgets \cite{chen2019automated}.

Some work focuses on the GUI widget understanding and the automation of the GUI testing \cite{feng2022gifdroid, liu2022guided, liu2022nighthawk}.
Chang \etal present a new approach that uses CV technologies to automate GUI testing and execute GUI actions \cite{chang2010gui}.
Nguyen \etal first introduce an approach, namely REMAUI \cite{nguyen2015reverse}, to use input images to identify GUI elements such as texts, images, and containers, using computer vision and optical character recognition (OCR) techniques. Chen \etal propose an approach that combines traditional CV technologies and DL-based object detection models to identify the widgets in the GUI screenshots \cite{chen2020object}.
Moran \etal propose a machine learning-based model to construct prototypes of GUIs for mobile apps \cite{moran2018machine}.
Fang \etal \cite{fang2023test} propose a comprehensive framework to generate structured and comprehensible test reports based on the structured bug model to analyze heterogeneous data from testing results, including GUI images, and a set of bug inconsistencies to build a novel bug taxonomy. 

Some novel papers are related to the usage of GUI widget intention understanding.
Xiao \etal propose IconIntent \cite{xiao2019iconintent}, a tool used to understand the intents of GUI widgets to identify sensitivity. IconIntent analyzes the meta files of GUI and extracts textual information, and widget images.
Also, Xi \etal present DeepIntent \cite{xi2019deepintent}, which is developed to focus on the GUI widgets and examine whether the intentions reflected in their GUIs justify their actual permission uses.
Zhang \etal propose a technique that creates accessibility metadata with an understanding of app GUI screenshots \cite{zhang2021screen}.
Chen \etal present an approach to automatically add labels to GUI components using deep learning models \cite{chen2020unblind}.
The above work inspires and supports us to analyze GUI images to extract the corresponding test operation intentions since GUI directly delivers app intentions to users.

\section{Conclusion}

Program understanding has long been a hot topic in the software engineering community, while few or even no studies have been done to help understand the test part (\ie test scripts) of an AUT. However, the intentions of test scripts have to be explored from the GUI information or the source code of AUTs, so existing technologies may lose effect when applied to test scripts. Mapping the test operations to corresponding source code or GUI elements is important. 

With the GUI-intensive and event-driven feature of mobile apps, \toolname provides an operation sequence model that formally represents test scripts, and \gui and \code work together in \toolname. \gui is used to process the GUI information of the operation, including the textual information and widget image intentions. \code explores the business logic contained in response methods, reflecting the test operation intentions. The results are aggregated and processed to generate comprehensible test script intentions. To evaluate the effectiveness of \toolname, we design an experiment attached with a user study. The results show that \toolname can effectively assist app developers in understanding test scripts and save their time.

\toolname can be viewed as a fundamental technique, as the source code comment generation techniques, and can have further applications for the test script maintenance and optimization. For example, with the test script intention, it is expected to assist the cross-app GUI test script recommendation to relieve the app developers' burden. Moreover, test script intentions generated by \toolname can be utilized to assist the test script slicing, which is also an important topic in white-box test case optimization, and the test script slicing can help better optimize the test scripts and locate the bugs.

With regard to the generalizability and scalability of \toolname, though our implementation is focused on the test script developed in Java and with Appium Driver because it is the mainstream for mobile app test script, we believe that \toolname can be easily adapted to other drivers and languages. The critical kernel of \toolname is to identify the identifiers of widgets in test scripts and corresponding widgets, and then map the identifiers to GUI image information and sourced code information. The GUI image and code understanding technologies can be reused in different test script drivers and languages.

In conclusion, \toolname is the first approach that helps understand GUI test script intentions and assists app developers in automatically generating user expectations of app behaviors for specific testing operations in test scripts (\ie intentions) and starts the automated assistance targeted at mobile app testing.

\begin{acks}
The authors would like to thank the editors and anonymous reviewers for their time and comments. This work is partially supported by the National Natural Science Foundation of China (U24A20337, 62272220, 62372228), and the Fundamental Research Funds for the Central Universities (14380029).
\end{acks}

\bibliographystyle{ACM-Reference-Format}
\bibliography{main}


\begin{thebibliography}{66}


\ifx \showCODEN    \undefined \def \showCODEN     #1{\unskip}     \fi
\ifx \showDOI      \undefined \def \showDOI       #1{#1}\fi
\ifx \showISBNx    \undefined \def \showISBNx     #1{\unskip}     \fi
\ifx \showISBNxiii \undefined \def \showISBNxiii  #1{\unskip}     \fi
\ifx \showISSN     \undefined \def \showISSN      #1{\unskip}     \fi
\ifx \showLCCN     \undefined \def \showLCCN      #1{\unskip}     \fi
\ifx \shownote     \undefined \def \shownote      #1{#1}          \fi
\ifx \showarticletitle \undefined \def \showarticletitle #1{#1}   \fi
\ifx \showURL      \undefined \def \showURL       {\relax}        \fi
\providecommand\bibfield[2]{#2}
\providecommand\bibinfo[2]{#2}
\providecommand\natexlab[1]{#1}
\providecommand\showeprint[2][]{arXiv:#2}

\bibitem[Allamanis et~al\mbox{.}(2015)]%
        {allamanis2015suggesting}
\bibfield{author}{\bibinfo{person}{Miltiadis Allamanis}, \bibinfo{person}{Earl~T. Barr}, \bibinfo{person}{Christian Bird}, {and} \bibinfo{person}{Charles Sutton}.} \bibinfo{year}{2015}\natexlab{}.
\newblock \showarticletitle{Suggesting accurate method and class names}. In \bibinfo{booktitle}{\emph{Proceedings of the 2015 10th Joint Meeting on Foundations of Software Engineering, 2015, Bergamo, Italy, August 30 - September 4, 2015}}. \bibinfo{publisher}{{ACM}}, \bibinfo{pages}{38--49}.
\newblock


\bibitem[Allamanis et~al\mbox{.}(2016)]%
        {allamanis2016convolutional}
\bibfield{author}{\bibinfo{person}{Miltiadis Allamanis}, \bibinfo{person}{Hao Peng}, {and} \bibinfo{person}{Charles Sutton}.} \bibinfo{year}{2016}\natexlab{}.
\newblock \showarticletitle{A convolutional attention network for extreme summarization of source code}. In \bibinfo{booktitle}{\emph{Proceedings of the 33rd International Conference on Machine Learning}}. \bibinfo{pages}{2091--2100}.
\newblock


\bibitem[Alon et~al\mbox{.}(2019a)]%
        {alon2018code2seq}
\bibfield{author}{\bibinfo{person}{Uri Alon}, \bibinfo{person}{Shaked Brody}, \bibinfo{person}{Omer Levy}, {and} \bibinfo{person}{Eran Yahav}.} \bibinfo{year}{2019}\natexlab{a}.
\newblock \showarticletitle{code2seq: Generating Sequences from Structured Representations of Code}. In \bibinfo{booktitle}{\emph{7th International Conference on Learning Representations, {ICLR} 2019, New Orleans, LA, USA, May 6-9, 2019}}. \bibinfo{publisher}{OpenReview.net}.
\newblock


\bibitem[Alon et~al\mbox{.}(2019b)]%
        {alon2019code2vec}
\bibfield{author}{\bibinfo{person}{Uri Alon}, \bibinfo{person}{Meital Zilberstein}, \bibinfo{person}{Omer Levy}, {and} \bibinfo{person}{Eran Yahav}.} \bibinfo{year}{2019}\natexlab{b}.
\newblock \showarticletitle{code2vec: Learning distributed representations of code}.
\newblock \bibinfo{journal}{\emph{Proceedings of the ACM on Programming Languages}} \bibinfo{volume}{3}, \bibinfo{number}{POPL} (\bibinfo{year}{2019}), \bibinfo{pages}{1--29}.
\newblock


\bibitem[Bahdanau et~al\mbox{.}(2014)]%
        {bahdanau2014neural}
\bibfield{author}{\bibinfo{person}{Dzmitry Bahdanau}, \bibinfo{person}{Kyunghyun Cho}, {and} \bibinfo{person}{Yoshua Bengio}.} \bibinfo{year}{2014}\natexlab{}.
\newblock \showarticletitle{Neural machine translation by jointly learning to align and translate}.
\newblock \bibinfo{journal}{\emph{arXiv preprint arXiv:1409.0473}} (\bibinfo{year}{2014}).
\newblock


\bibitem[Banerjee and Lavie(2005)]%
        {banerjee2005meteor}
\bibfield{author}{\bibinfo{person}{Satanjeev Banerjee} {and} \bibinfo{person}{Alon Lavie}.} \bibinfo{year}{2005}\natexlab{}.
\newblock \showarticletitle{METEOR: An automatic metric for MT evaluation with improved correlation with human judgments}. In \bibinfo{booktitle}{\emph{Proceedings of the ACL Workshop on Intrinsic and Extrinsic Evaluation Measures for Machine Rranslation and/or Summarization}}. \bibinfo{pages}{65--72}.
\newblock


\bibitem[Beltramelli(2018)]%
        {beltramelli2018pix2code}
\bibfield{author}{\bibinfo{person}{Tony Beltramelli}.} \bibinfo{year}{2018}\natexlab{}.
\newblock \showarticletitle{pix2code: Generating code from a graphical user interface screenshot}. In \bibinfo{booktitle}{\emph{Proceedings of the ACM SIGCHI Symposium on Engineering Interactive Computing Systems}}. \bibinfo{pages}{1--6}.
\newblock


\bibitem[Chang et~al\mbox{.}(2010)]%
        {chang2010gui}
\bibfield{author}{\bibinfo{person}{Tsung-Hsiang Chang}, \bibinfo{person}{Tom Yeh}, {and} \bibinfo{person}{Robert~C Miller}.} \bibinfo{year}{2010}\natexlab{}.
\newblock \showarticletitle{GUI testing using computer vision}. In \bibinfo{booktitle}{\emph{Proceedings of the SIGCHI Conference on Human Factors in Computing Systems}}. \bibinfo{pages}{1535--1544}.
\newblock


\bibitem[Chen et~al\mbox{.}(2018)]%
        {chen2018ui}
\bibfield{author}{\bibinfo{person}{Chunyang Chen}, \bibinfo{person}{Ting Su}, \bibinfo{person}{Guozhu Meng}, \bibinfo{person}{Zhenchang Xing}, {and} \bibinfo{person}{Yang Liu}.} \bibinfo{year}{2018}\natexlab{}.
\newblock \showarticletitle{From UI design image to GUI skeleton: a neural machine translator to bootstrap mobile GUI implementation}. In \bibinfo{booktitle}{\emph{Proceedings of the 40th International Conference on Software Engineering}}. \bibinfo{pages}{665--676}.
\newblock


\bibitem[Chen et~al\mbox{.}(2020a)]%
        {chen2020unblind}
\bibfield{author}{\bibinfo{person}{Jieshan Chen}, \bibinfo{person}{Chunyang Chen}, \bibinfo{person}{Zhenchang Xing}, \bibinfo{person}{Xiwei Xu}, \bibinfo{person}{Liming Zhu}, \bibinfo{person}{Guoqiang Li}, {and} \bibinfo{person}{Jinshui Wang}.} \bibinfo{year}{2020}\natexlab{a}.
\newblock \showarticletitle{Unblind your apps: Predicting natural-language labels for mobile gui components by deep learning}. In \bibinfo{booktitle}{\emph{Proceedings of the 42nd IEEE/ACM International Conference on Software Engineering}}. IEEE, \bibinfo{pages}{322--334}.
\newblock


\bibitem[Chen et~al\mbox{.}(2020b)]%
        {chen2020object}
\bibfield{author}{\bibinfo{person}{Jieshan Chen}, \bibinfo{person}{Mulong Xie}, \bibinfo{person}{Zhenchang Xing}, \bibinfo{person}{Chunyang Chen}, \bibinfo{person}{Xiwei Xu}, \bibinfo{person}{Liming Zhu}, {and} \bibinfo{person}{Guoqiang Li}.} \bibinfo{year}{2020}\natexlab{b}.
\newblock \showarticletitle{Object detection for graphical user interface: old fashioned or deep learning or a combination?}. In \bibinfo{booktitle}{\emph{Proceedings of the 28th ACM Joint Meeting on European Software Engineering Conference and Symposium on the Foundations of Software Engineering}}. \bibinfo{pages}{1202--1214}.
\newblock


\bibitem[Chen et~al\mbox{.}(2019)]%
        {chen2019automated}
\bibfield{author}{\bibinfo{person}{Sen Chen}, \bibinfo{person}{Lingling Fan}, \bibinfo{person}{Ting Su}, \bibinfo{person}{Lei Ma}, \bibinfo{person}{Yang Liu}, {and} \bibinfo{person}{Lihua Xu}.} \bibinfo{year}{2019}\natexlab{}.
\newblock \showarticletitle{Automated cross-platform GUI code generation for mobile apps}. In \bibinfo{booktitle}{\emph{Proceedings of the 1st IEEE International Workshop on Artificial Intelligence for Mobile}}. IEEE, \bibinfo{pages}{13--16}.
\newblock


\bibitem[Daka et~al\mbox{.}(2017)]%
        {daka2017generating}
\bibfield{author}{\bibinfo{person}{Ermira Daka}, \bibinfo{person}{Jos{\'e}~Miguel Rojas}, {and} \bibinfo{person}{Gordon Fraser}.} \bibinfo{year}{2017}\natexlab{}.
\newblock \showarticletitle{Generating unit tests with descriptive names or: Would you name your children thing1 and thing2?}. In \bibinfo{booktitle}{\emph{Proceedings of the 26th ACM SIGSOFT International Symposium on Software Testing and Analysis}}. \bibinfo{pages}{57--67}.
\newblock


\bibitem[Fang et~al\mbox{.}(2023)]%
        {fang2023test}
\bibfield{author}{\bibinfo{person}{Chunrong Fang}, \bibinfo{person}{Shengcheng Yu}, \bibinfo{person}{Ting Su}, \bibinfo{person}{Jing Zhang}, \bibinfo{person}{Yuanhan Tian}, {and} \bibinfo{person}{Yang Liu}.} \bibinfo{year}{2023}\natexlab{}.
\newblock \showarticletitle{Test Report Generation for Android App Testing via Heterogeneous Data Analysis}.
\newblock \bibinfo{journal}{\emph{IEEE Transactions on Software Engineering}} (\bibinfo{year}{2023}).
\newblock


\bibitem[Fang et~al\mbox{.}(2022)]%
        {fang2022prhan}
\bibfield{author}{\bibinfo{person}{Sen Fang}, \bibinfo{person}{Tao Zhang}, \bibinfo{person}{You-Shuai Tan}, \bibinfo{person}{Zhou Xu}, \bibinfo{person}{Zhi-Xin Yuan}, {and} \bibinfo{person}{Ling-Ze Meng}.} \bibinfo{year}{2022}\natexlab{}.
\newblock \showarticletitle{PRHAN: Automated Pull Request Description Generation Based on Hybrid Attention Network}.
\newblock \bibinfo{journal}{\emph{Journal of Systems and Software}}  \bibinfo{volume}{185} (\bibinfo{year}{2022}), \bibinfo{pages}{111160}.
\newblock


\bibitem[Feng and Chen(2022)]%
        {feng2022gifdroid}
\bibfield{author}{\bibinfo{person}{Sidong Feng} {and} \bibinfo{person}{Chunyang Chen}.} \bibinfo{year}{2022}\natexlab{}.
\newblock \showarticletitle{GIFdroid: Automated Replay of Visual Bug Reports for Android Apps}. In \bibinfo{booktitle}{\emph{Proceedings of the 43rd IEEE/ACM International Conference on Software Engineering}}. IEEE.
\newblock


\bibitem[Gao et~al\mbox{.}(2023)]%
        {gao2023code}
\bibfield{author}{\bibinfo{person}{Shuzheng Gao}, \bibinfo{person}{Cuiyun Gao}, \bibinfo{person}{Yulan He}, \bibinfo{person}{Jichuan Zeng}, \bibinfo{person}{Lunyiu Nie}, \bibinfo{person}{Xin Xia}, {and} \bibinfo{person}{Michael Lyu}.} \bibinfo{year}{2023}\natexlab{}.
\newblock \showarticletitle{Code Structure--Guided Transformer for Source Code Summarization}.
\newblock \bibinfo{journal}{\emph{ACM Transactions on Software Engineering and Methodology}} \bibinfo{volume}{32}, \bibinfo{number}{1} (\bibinfo{year}{2023}), \bibinfo{pages}{1--32}.
\newblock


\bibitem[Geng et~al\mbox{.}(2024)]%
        {geng2024large}
\bibfield{author}{\bibinfo{person}{Mingyang Geng}, \bibinfo{person}{Shangwen Wang}, \bibinfo{person}{Dezun Dong}, \bibinfo{person}{Haotian Wang}, \bibinfo{person}{Ge Li}, \bibinfo{person}{Zhi Jin}, \bibinfo{person}{Xiaoguang Mao}, {and} \bibinfo{person}{Xiangke Liao}.} \bibinfo{year}{2024}\natexlab{}.
\newblock \showarticletitle{Large Language Models are Few-Shot Summarizers: Multi-Intent Comment Generation via In-Context Learning}. In \bibinfo{booktitle}{\emph{Proceedings of the IEEE/ACM 46th International Conference on Software Engineering}}. Article \bibinfo{articleno}{39}, \bibinfo{numpages}{13}~pages.
\newblock


\bibitem[Haiduc et~al\mbox{.}(2010a)]%
        {haiduc2010supporting}
\bibfield{author}{\bibinfo{person}{Sonia Haiduc}, \bibinfo{person}{Jairo Aponte}, {and} \bibinfo{person}{Andrian Marcus}.} \bibinfo{year}{2010}\natexlab{a}.
\newblock \showarticletitle{Supporting program comprehension with source code summarization}. In \bibinfo{booktitle}{\emph{Proceedings of the 32nd ACM/IEEE International Conference on Software Engineering}}, Vol.~\bibinfo{volume}{2}. IEEE, \bibinfo{pages}{223--226}.
\newblock


\bibitem[Haiduc et~al\mbox{.}(2010b)]%
        {haiduc2010use}
\bibfield{author}{\bibinfo{person}{Sonia Haiduc}, \bibinfo{person}{Jairo Aponte}, \bibinfo{person}{Laura Moreno}, {and} \bibinfo{person}{Andrian Marcus}.} \bibinfo{year}{2010}\natexlab{b}.
\newblock \showarticletitle{On the use of automated text summarization techniques for summarizing source code}. In \bibinfo{booktitle}{\emph{Proceedings of the 17th Working Conference on Reverse Engineering}}. IEEE, \bibinfo{pages}{35--44}.
\newblock


\bibitem[Haije et~al\mbox{.}(2016)]%
        {haije2016automatic}
\bibfield{author}{\bibinfo{person}{Tjalling Haije}, \bibinfo{person}{Bachelor Opleiding~Kunstmatige Intelligentie}, \bibinfo{person}{E Gavves}, {and} \bibinfo{person}{H Heuer}.} \bibinfo{year}{2016}\natexlab{}.
\newblock \showarticletitle{Automatic comment generation using a neural translation model}.
\newblock \bibinfo{journal}{\emph{Information and Software Technology}} \bibinfo{volume}{55}, \bibinfo{number}{3} (\bibinfo{year}{2016}), \bibinfo{pages}{258--268}.
\newblock


\bibitem[Harman et~al\mbox{.}(2016)]%
        {harman2016mobile}
\bibfield{author}{\bibinfo{person}{Mark Harman}, \bibinfo{person}{Afnan Al-Subaihin}, \bibinfo{person}{Yue Jia}, \bibinfo{person}{William Martin}, \bibinfo{person}{Federica Sarro}, {and} \bibinfo{person}{Yuanyuan Zhang}.} \bibinfo{year}{2016}\natexlab{}.
\newblock \showarticletitle{Mobile app and app store analysis, testing and optimisation}. In \bibinfo{booktitle}{\emph{Proceedings of the International Conference on Mobile Software Engineering and Systems}}. \bibinfo{pages}{243--244}.
\newblock


\bibitem[Hochreiter and Schmidhuber(1997)]%
        {hochreiter1997long}
\bibfield{author}{\bibinfo{person}{Sepp Hochreiter} {and} \bibinfo{person}{J{\"u}rgen Schmidhuber}.} \bibinfo{year}{1997}\natexlab{}.
\newblock \showarticletitle{Long short-term memory}.
\newblock \bibinfo{journal}{\emph{Neural computation}} \bibinfo{volume}{9}, \bibinfo{number}{8} (\bibinfo{year}{1997}), \bibinfo{pages}{1735--1780}.
\newblock


\bibitem[Hu et~al\mbox{.}(2018)]%
        {hu2018deep}
\bibfield{author}{\bibinfo{person}{Xing Hu}, \bibinfo{person}{Ge Li}, \bibinfo{person}{Xin Xia}, \bibinfo{person}{David Lo}, {and} \bibinfo{person}{Zhi Jin}.} \bibinfo{year}{2018}\natexlab{}.
\newblock \showarticletitle{Deep code comment generation}. In \bibinfo{booktitle}{\emph{Proceedings of the 26th IEEE/ACM International Conference on Program Comprehension}}. IEEE, \bibinfo{pages}{200--20010}.
\newblock


\bibitem[Hu et~al\mbox{.}(2020)]%
        {hu2020deep}
\bibfield{author}{\bibinfo{person}{Xing Hu}, \bibinfo{person}{Ge Li}, \bibinfo{person}{Xin Xia}, \bibinfo{person}{David Lo}, {and} \bibinfo{person}{Zhi Jin}.} \bibinfo{year}{2020}\natexlab{}.
\newblock \showarticletitle{Deep code comment generation with hybrid lexical and syntactical information}.
\newblock \bibinfo{journal}{\emph{Empirical Software Engineering}} \bibinfo{volume}{25}, \bibinfo{number}{3} (\bibinfo{year}{2020}), \bibinfo{pages}{2179--2217}.
\newblock


\bibitem[Huang et~al\mbox{.}(2023)]%
        {huang2023bcgen}
\bibfield{author}{\bibinfo{person}{Yuan Huang}, \bibinfo{person}{Jinbo Huang}, \bibinfo{person}{Xiangping Chen}, \bibinfo{person}{Kunning He}, {and} \bibinfo{person}{Xiaocong Zhou}.} \bibinfo{year}{2023}\natexlab{}.
\newblock \showarticletitle{BCGen: a comment generation method for bytecode}.
\newblock \bibinfo{journal}{\emph{Automated Software Engineering}} \bibinfo{volume}{30}, \bibinfo{number}{1} (\bibinfo{year}{2023}), \bibinfo{pages}{5}.
\newblock


\bibitem[Iyer et~al\mbox{.}(2016)]%
        {iyer2016summarizing}
\bibfield{author}{\bibinfo{person}{Srinivasan Iyer}, \bibinfo{person}{Ioannis Konstas}, \bibinfo{person}{Alvin Cheung}, {and} \bibinfo{person}{Luke Zettlemoyer}.} \bibinfo{year}{2016}\natexlab{}.
\newblock \showarticletitle{Summarizing source code using a neural attention model}. In \bibinfo{booktitle}{\emph{Proceedings of the 54th Annual Meeting of the Association for Computational Linguistics}}. \bibinfo{pages}{2073--2083}.
\newblock


\bibitem[Jang et~al\mbox{.}(2020)]%
        {jang2020approach}
\bibfield{author}{\bibinfo{person}{Seunghui Jang}, \bibinfo{person}{Ki~Yong Lee}, {and} \bibinfo{person}{Yanggon Kim}.} \bibinfo{year}{2020}\natexlab{}.
\newblock \showarticletitle{An Approach to Improving the Effectiveness of Data Augmentation for Deep Neural Networks}. In \bibinfo{booktitle}{\emph{Proceedings of the 44th IEEE Annual Computers, Software, and Applications Conference}}. IEEE, \bibinfo{pages}{1290--1295}.
\newblock


\bibitem[LeClair et~al\mbox{.}(2019)]%
        {leclair2019neural}
\bibfield{author}{\bibinfo{person}{Alexander LeClair}, \bibinfo{person}{Siyuan Jiang}, {and} \bibinfo{person}{Collin McMillan}.} \bibinfo{year}{2019}\natexlab{}.
\newblock \showarticletitle{A neural model for generating natural language summaries of program subroutines}. In \bibinfo{booktitle}{\emph{Proceedings of the 41st IEEE/ACM International Conference on Software Engineering}}. IEEE.
\newblock


\bibitem[Li et~al\mbox{.}(2018)]%
        {li2018aiding}
\bibfield{author}{\bibinfo{person}{Boyang Li}, \bibinfo{person}{Christopher Vendome}, \bibinfo{person}{Mario Linares-V{\'a}squez}, {and} \bibinfo{person}{Denys Poshyvanyk}.} \bibinfo{year}{2018}\natexlab{}.
\newblock \showarticletitle{Aiding comprehension of unit test cases and test suites with stereotype-based tagging}. In \bibinfo{booktitle}{\emph{Proceedings of the 26th Conference on Program Comprehension}}. \bibinfo{pages}{52--63}.
\newblock


\bibitem[Li et~al\mbox{.}(2016)]%
        {li2016automatically}
\bibfield{author}{\bibinfo{person}{Boyang Li}, \bibinfo{person}{Christopher Vendome}, \bibinfo{person}{Mario Linares-V{\'a}squez}, \bibinfo{person}{Denys Poshyvanyk}, {and} \bibinfo{person}{Nicholas~A Kraft}.} \bibinfo{year}{2016}\natexlab{}.
\newblock \showarticletitle{Automatically documenting unit test cases}. In \bibinfo{booktitle}{\emph{Proceedings of the IEEE International Conference on Software Testing, Verification and Validation}}. IEEE, \bibinfo{pages}{341--352}.
\newblock


\bibitem[Li et~al\mbox{.}(2019)]%
        {li2019boosting}
\bibfield{author}{\bibinfo{person}{Zenan Li}, \bibinfo{person}{Xiaoxing Ma}, \bibinfo{person}{Chang Xu}, \bibinfo{person}{Chun Cao}, \bibinfo{person}{Jingwei Xu}, {and} \bibinfo{person}{Jian L{\"u}}.} \bibinfo{year}{2019}\natexlab{}.
\newblock \showarticletitle{Boosting operational dnn testing efficiency through conditioning}. In \bibinfo{booktitle}{\emph{Proceedings of the 27th ACM Joint Meeting on European Software Engineering Conference and Symposium on the Foundations of Software Engineering}}. \bibinfo{pages}{499--509}.
\newblock


\bibitem[Liang and Zhu(2018)]%
        {liang2018automatic}
\bibfield{author}{\bibinfo{person}{Yuding Liang} {and} \bibinfo{person}{Kenny Zhu}.} \bibinfo{year}{2018}\natexlab{}.
\newblock \showarticletitle{Automatic generation of text descriptive comments for code blocks}. In \bibinfo{booktitle}{\emph{Proceedings of the 32nd AAAI Conference on Artificial Intelligence}}, Vol.~\bibinfo{volume}{32}.
\newblock


\bibitem[Lin and Hovy(2003)]%
        {lin2003automatic}
\bibfield{author}{\bibinfo{person}{Chin-Yew Lin} {and} \bibinfo{person}{Eduard Hovy}.} \bibinfo{year}{2003}\natexlab{}.
\newblock \showarticletitle{Automatic evaluation of summaries using n-gram co-occurrence statistics}. In \bibinfo{booktitle}{\emph{Proceedings of the Human Language Technology Conference of the North American Chapter of the Association for Computational Linguistics}}. \bibinfo{pages}{150--157}.
\newblock


\bibitem[Liu et~al\mbox{.}(2020)]%
        {liu2020androzooopen}
\bibfield{author}{\bibinfo{person}{Pei Liu}, \bibinfo{person}{Li Li}, \bibinfo{person}{Yanjie Zhao}, \bibinfo{person}{Xiaoyu Sun}, {and} \bibinfo{person}{John Grundy}.} \bibinfo{year}{2020}\natexlab{}.
\newblock \showarticletitle{AndroZooOpen: Collecting Large-scale Open Source Android Apps for the Research Community}. In \bibinfo{booktitle}{\emph{Proceedings of the 17th International Conference on Mining Software Repositories}}. \bibinfo{pages}{548--552}.
\newblock


\bibitem[Liu et~al\mbox{.}(2022a)]%
        {liu2022guided}
\bibfield{author}{\bibinfo{person}{Zhe Liu}, \bibinfo{person}{Chunyang Chen}, \bibinfo{person}{Junjie Wang}, \bibinfo{person}{Yuekai Huang}, \bibinfo{person}{Jun Hu}, {and} \bibinfo{person}{Qing Wang}.} \bibinfo{year}{2022}\natexlab{a}.
\newblock \showarticletitle{Guided Bug Crush: Assist Manual GUI Testing of Android Apps via Hint Moves}. In \bibinfo{booktitle}{\emph{Proceedings of the CHI Conference on Human Factors in Computing Systems}}.
\newblock


\bibitem[Liu et~al\mbox{.}(2022b)]%
        {liu2022nighthawk}
\bibfield{author}{\bibinfo{person}{Zhe Liu}, \bibinfo{person}{Chunyang Chen}, \bibinfo{person}{Junjie Wang}, \bibinfo{person}{Yuekai Huang}, \bibinfo{person}{Jun Hu}, {and} \bibinfo{person}{Qing Wang}.} \bibinfo{year}{2022}\natexlab{b}.
\newblock \showarticletitle{Nighthawk: Fully Automated Localizing UI Display Issues via Visual Understanding}.
\newblock \bibinfo{journal}{\emph{IEEE Transactions on Software Engineering}} (\bibinfo{year}{2022}).
\newblock


\bibitem[Malhotra and Chhabra(2018)]%
        {malhotra2018class}
\bibfield{author}{\bibinfo{person}{Mrinaal Malhotra} {and} \bibinfo{person}{Jitender~Kumar Chhabra}.} \bibinfo{year}{2018}\natexlab{}.
\newblock \showarticletitle{Class level code summarization based on dependencies and micro patterns}. In \bibinfo{booktitle}{\emph{Proceedings of the 2nd International Conference on Inventive Communication and Computational Technologies}}. IEEE, \bibinfo{pages}{1011--1016}.
\newblock


\bibitem[Mikolov et~al\mbox{.}(2013)]%
        {mikolov2013distributed}
\bibfield{author}{\bibinfo{person}{Tomas Mikolov}, \bibinfo{person}{Ilya Sutskever}, \bibinfo{person}{Kai Chen}, \bibinfo{person}{Greg~S Corrado}, {and} \bibinfo{person}{Jeff Dean}.} \bibinfo{year}{2013}\natexlab{}.
\newblock \showarticletitle{Distributed representations of words and phrases and their compositionality}. In \bibinfo{booktitle}{\emph{Advances in neural information processing systems}}. \bibinfo{pages}{3111--3119}.
\newblock


\bibitem[Moran et~al\mbox{.}(2018)]%
        {moran2018machine}
\bibfield{author}{\bibinfo{person}{Kevin Moran}, \bibinfo{person}{Carlos Bernal-C{\'a}rdenas}, \bibinfo{person}{Michael Curcio}, \bibinfo{person}{Richard Bonett}, {and} \bibinfo{person}{Denys Poshyvanyk}.} \bibinfo{year}{2018}\natexlab{}.
\newblock \showarticletitle{Machine learning-based prototyping of graphical user interfaces for mobile apps}.
\newblock \bibinfo{journal}{\emph{IEEE Transactions on Software Engineering}} \bibinfo{volume}{46}, \bibinfo{number}{2} (\bibinfo{year}{2018}), \bibinfo{pages}{196--221}.
\newblock


\bibitem[Moreno et~al\mbox{.}(2013)]%
        {moreno2013automatic}
\bibfield{author}{\bibinfo{person}{Laura Moreno}, \bibinfo{person}{Jairo Aponte}, \bibinfo{person}{Giriprasad Sridhara}, \bibinfo{person}{Andrian Marcus}, \bibinfo{person}{Lori Pollock}, {and} \bibinfo{person}{K Vijay-Shanker}.} \bibinfo{year}{2013}\natexlab{}.
\newblock \showarticletitle{Automatic generation of natural language summaries for java classes}. In \bibinfo{booktitle}{\emph{Proceedings of the 21st International Conference on Program Comprehension}}. IEEE, \bibinfo{pages}{23--32}.
\newblock


\bibitem[Mu et~al\mbox{.}(2023)]%
        {mu2023developer}
\bibfield{author}{\bibinfo{person}{Fangwen Mu}, \bibinfo{person}{Xiao Chen}, \bibinfo{person}{Lin Shi}, \bibinfo{person}{Song Wang}, {and} \bibinfo{person}{Qing Wang}.} \bibinfo{year}{2023}\natexlab{}.
\newblock \showarticletitle{Developer-intent driven code comment generation}. In \bibinfo{booktitle}{\emph{IEEE/ACM 45th International Conference on Software Engineering}}. IEEE, \bibinfo{pages}{768--780}.
\newblock


\bibitem[Nazar et~al\mbox{.}(2016)]%
        {nazar2016source}
\bibfield{author}{\bibinfo{person}{Najam Nazar}, \bibinfo{person}{He Jiang}, \bibinfo{person}{Guojun Gao}, \bibinfo{person}{Tao Zhang}, \bibinfo{person}{Xiaochen Li}, {and} \bibinfo{person}{Zhilei Ren}.} \bibinfo{year}{2016}\natexlab{}.
\newblock \showarticletitle{Source code fragment summarization with small-scale crowdsourcing based features}.
\newblock \bibinfo{journal}{\emph{Frontiers of Computer Science}} \bibinfo{volume}{10}, \bibinfo{number}{3} (\bibinfo{year}{2016}), \bibinfo{pages}{504--517}.
\newblock


\bibitem[Nguyen and Csallner(2015)]%
        {nguyen2015reverse}
\bibfield{author}{\bibinfo{person}{Tuan~Anh Nguyen} {and} \bibinfo{person}{Christoph Csallner}.} \bibinfo{year}{2015}\natexlab{}.
\newblock \showarticletitle{Reverse engineering mobile application user interfaces with remaui (t)}. In \bibinfo{booktitle}{\emph{Proceedings of the 30th IEEE/ACM International Conference on Automated Software Engineering}}. IEEE, \bibinfo{pages}{248--259}.
\newblock


\bibitem[Panichella et~al\mbox{.}(2016)]%
        {panichella2016impact}
\bibfield{author}{\bibinfo{person}{Sebastiano Panichella}, \bibinfo{person}{Annibale Panichella}, \bibinfo{person}{Moritz Beller}, \bibinfo{person}{Andy Zaidman}, {and} \bibinfo{person}{Harald~C Gall}.} \bibinfo{year}{2016}\natexlab{}.
\newblock \showarticletitle{The impact of test case summaries on bug fixing performance: An empirical investigation}. In \bibinfo{booktitle}{\emph{Proceedings of the 38th International Conference on Software Engineering}}. \bibinfo{pages}{547--558}.
\newblock


\bibitem[Papineni et~al\mbox{.}(2002)]%
        {papineni2002bleu}
\bibfield{author}{\bibinfo{person}{Kishore Papineni}, \bibinfo{person}{Salim Roukos}, \bibinfo{person}{Todd Ward}, {and} \bibinfo{person}{Wei-Jing Zhu}.} \bibinfo{year}{2002}\natexlab{}.
\newblock \showarticletitle{Bleu: a method for automatic evaluation of machine translation}. In \bibinfo{booktitle}{\emph{Proceedings of the 40th Annual Meeting of the Association for Computational Linguistics}}. \bibinfo{pages}{311--318}.
\newblock


\bibitem[Rai et~al\mbox{.}(2017)]%
        {rai2017method}
\bibfield{author}{\bibinfo{person}{Sawan Rai}, \bibinfo{person}{Tejaswini Gaikwad}, \bibinfo{person}{Sparshi Jain}, {and} \bibinfo{person}{Atul Gupta}.} \bibinfo{year}{2017}\natexlab{}.
\newblock \showarticletitle{Method level text summarization for java code using nano-patterns}. In \bibinfo{booktitle}{\emph{Proceedings of the 24th Asia-Pacific Software Engineering Conference}}. IEEE, \bibinfo{pages}{199--208}.
\newblock


\bibitem[Rodeghero et~al\mbox{.}(2015)]%
        {rodeghero2015eye}
\bibfield{author}{\bibinfo{person}{Paige Rodeghero}, \bibinfo{person}{Cheng Liu}, \bibinfo{person}{Paul~W McBurney}, {and} \bibinfo{person}{Collin McMillan}.} \bibinfo{year}{2015}\natexlab{}.
\newblock \showarticletitle{An eye-tracking study of java programmers and application to source code summarization}.
\newblock \bibinfo{journal}{\emph{IEEE Transactions on Software Engineering}} (\bibinfo{year}{2015}), \bibinfo{pages}{1038--1054}.
\newblock


\bibitem[Rodeghero et~al\mbox{.}(2014)]%
        {rodeghero2014improving}
\bibfield{author}{\bibinfo{person}{Paige Rodeghero}, \bibinfo{person}{Collin McMillan}, \bibinfo{person}{Paul~W McBurney}, \bibinfo{person}{Nigel Bosch}, {and} \bibinfo{person}{Sidney D'Mello}.} \bibinfo{year}{2014}\natexlab{}.
\newblock \showarticletitle{Improving automated source code summarization via an eye-tracking study of programmers}. In \bibinfo{booktitle}{\emph{Proceedings of the 36th International Conference on Software Engineering}}. \bibinfo{pages}{390--401}.
\newblock


\bibitem[Roy et~al\mbox{.}(2020)]%
        {roy2020deeptc}
\bibfield{author}{\bibinfo{person}{Devjeet Roy}, \bibinfo{person}{Ziyi Zhang}, \bibinfo{person}{Maggie Ma}, \bibinfo{person}{Venera Arnaoudova}, \bibinfo{person}{Annibale Panichella}, \bibinfo{person}{Sebastiano Panichella}, \bibinfo{person}{Danielle Gonzalez}, {and} \bibinfo{person}{Mehdi Mirakhorli}.} \bibinfo{year}{2020}\natexlab{}.
\newblock \showarticletitle{DeepTC-Enhancer: Improving the readability of automatically generated tests}. In \bibinfo{booktitle}{\emph{Proceedings of the 35th IEEE/ACM International Conference on Automated Software Engineering}}. IEEE, \bibinfo{pages}{287--298}.
\newblock


\bibitem[Salman et~al\mbox{.}(2015)]%
        {salman2015students}
\bibfield{author}{\bibinfo{person}{Iflaah Salman}, \bibinfo{person}{Ayse~Tosun Misirli}, {and} \bibinfo{person}{Natalia Juristo}.} \bibinfo{year}{2015}\natexlab{}.
\newblock \showarticletitle{Are Students Representatives of Professionals in Software Engineering Experiments?}. In \bibinfo{booktitle}{\emph{Proceedings of the 37th IEEE International Conference on Software Engineering}}. IEEE, \bibinfo{pages}{666--676}.
\newblock


\bibitem[Simonyan and Zisserman(2014)]%
        {simonyan2014very}
\bibfield{author}{\bibinfo{person}{Karen Simonyan} {and} \bibinfo{person}{Andrew Zisserman}.} \bibinfo{year}{2014}\natexlab{}.
\newblock \showarticletitle{Very deep convolutional networks for large-scale image recognition}.
\newblock \bibinfo{journal}{\emph{arXiv preprint arXiv:1409.1556}} (\bibinfo{year}{2014}).
\newblock


\bibitem[Song et~al\mbox{.}(2019)]%
        {song2019survey}
\bibfield{author}{\bibinfo{person}{Xiaotao Song}, \bibinfo{person}{Hailong Sun}, \bibinfo{person}{Xu Wang}, {and} \bibinfo{person}{Jiafei Yan}.} \bibinfo{year}{2019}\natexlab{}.
\newblock \showarticletitle{A survey of automatic generation of source code comments: Algorithms and techniques}.
\newblock \bibinfo{journal}{\emph{IEEE Access}}  \bibinfo{volume}{7} (\bibinfo{year}{2019}), \bibinfo{pages}{111411--111428}.
\newblock


\bibitem[Sridhara et~al\mbox{.}(2010)]%
        {sridhara2010towards}
\bibfield{author}{\bibinfo{person}{Giriprasad Sridhara}, \bibinfo{person}{Emily Hill}, \bibinfo{person}{Divya Muppaneni}, \bibinfo{person}{Lori Pollock}, {and} \bibinfo{person}{K Vijay-Shanker}.} \bibinfo{year}{2010}\natexlab{}.
\newblock \showarticletitle{Towards automatically generating summary comments for java methods}. In \bibinfo{booktitle}{\emph{Proceedings of the 25th IEEE/ACM International Conference on Automated Software Engineering}}. \bibinfo{pages}{43--52}.
\newblock


\bibitem[Sridhara et~al\mbox{.}(2011)]%
        {sridhara2011automatically}
\bibfield{author}{\bibinfo{person}{Giriprasad Sridhara}, \bibinfo{person}{Lori Pollock}, {and} \bibinfo{person}{K Vijay-Shanker}.} \bibinfo{year}{2011}\natexlab{}.
\newblock \showarticletitle{Automatically detecting and describing high level actions within methods}. In \bibinfo{booktitle}{\emph{Proceedings of the 33rd International Conference on Software Engineering}}. IEEE, \bibinfo{pages}{101--110}.
\newblock


\bibitem[Vedantam et~al\mbox{.}(2015)]%
        {vedantam2015cider}
\bibfield{author}{\bibinfo{person}{Ramakrishna Vedantam}, \bibinfo{person}{C Lawrence~Zitnick}, {and} \bibinfo{person}{Devi Parikh}.} \bibinfo{year}{2015}\natexlab{}.
\newblock \showarticletitle{Cider: Consensus-based image description evaluation}. In \bibinfo{booktitle}{\emph{Proceedings of the IEEE Conference on Computer Vision and Pattern Recognition}}. \bibinfo{pages}{4566--4575}.
\newblock


\bibitem[Wong et~al\mbox{.}(2013)]%
        {wong2013autocomment}
\bibfield{author}{\bibinfo{person}{Edmund Wong}, \bibinfo{person}{Jinqiu Yang}, {and} \bibinfo{person}{Lin Tan}.} \bibinfo{year}{2013}\natexlab{}.
\newblock \showarticletitle{Autocomment: Mining question and answer sites for automatic comment generation}. In \bibinfo{booktitle}{\emph{Proceedings of the 28th IEEE/ACM International Conference on Automated Software Engineering}}. \bibinfo{pages}{562--567}.
\newblock


\bibitem[Xi et~al\mbox{.}(2019)]%
        {xi2019deepintent}
\bibfield{author}{\bibinfo{person}{Shengqu Xi}, \bibinfo{person}{Shao Yang}, \bibinfo{person}{Xusheng Xiao}, \bibinfo{person}{Yuan Yao}, \bibinfo{person}{Yayuan Xiong}, \bibinfo{person}{Fengyuan Xu}, \bibinfo{person}{Haoyu Wang}, \bibinfo{person}{Peng Gao}, \bibinfo{person}{Zhuotao Liu}, \bibinfo{person}{Feng Xu}, {et~al\mbox{.}}} \bibinfo{year}{2019}\natexlab{}.
\newblock \showarticletitle{DeepIntent: Deep icon-behavior learning for detecting intention-behavior discrepancy in mobile apps}. In \bibinfo{booktitle}{\emph{Proceedings of the ACM SIGSAC Conference on Computer and Communications Security}}. \bibinfo{pages}{2421--2436}.
\newblock


\bibitem[Xia et~al\mbox{.}(2017)]%
        {xia2017measuring}
\bibfield{author}{\bibinfo{person}{Xin Xia}, \bibinfo{person}{Lingfeng Bao}, \bibinfo{person}{David Lo}, \bibinfo{person}{Zhenchang Xing}, \bibinfo{person}{Ahmed~E Hassan}, {and} \bibinfo{person}{Shanping Li}.} \bibinfo{year}{2017}\natexlab{}.
\newblock \showarticletitle{Measuring program comprehension: A large-scale field study with professionals}.
\newblock \bibinfo{journal}{\emph{IEEE Transactions on Software Engineering}} \bibinfo{volume}{44}, \bibinfo{number}{10} (\bibinfo{year}{2017}), \bibinfo{pages}{951--976}.
\newblock


\bibitem[Xiao et~al\mbox{.}(2019)]%
        {xiao2019iconintent}
\bibfield{author}{\bibinfo{person}{Xusheng Xiao}, \bibinfo{person}{Xiaoyin Wang}, \bibinfo{person}{Zhihao Cao}, \bibinfo{person}{Hanlin Wang}, {and} \bibinfo{person}{Peng Gao}.} \bibinfo{year}{2019}\natexlab{}.
\newblock \showarticletitle{Iconintent: automatic identification of sensitive ui widgets based on icon classification for android apps}. In \bibinfo{booktitle}{\emph{Proceedings of the 41st IEEE/ACM International Conference on Software Engineering}}. IEEE, \bibinfo{pages}{257--268}.
\newblock


\bibitem[Yu et~al\mbox{.}(2021)]%
        {yu2021layout}
\bibfield{author}{\bibinfo{person}{Shengcheng Yu}, \bibinfo{person}{Chunrong Fang}, \bibinfo{person}{Yexiao Yun}, {and} \bibinfo{person}{Yang Feng}.} \bibinfo{year}{2021}\natexlab{}.
\newblock \showarticletitle{Layout and image recognition driving cross-platform automated mobile testing}. In \bibinfo{booktitle}{\emph{Proceedings of the 43rd IEEE/ACM International Conference on Software Engineering}}. IEEE, \bibinfo{pages}{1561--1571}.
\newblock


\bibitem[Zhang et~al\mbox{.}(2016)]%
        {zhang2016towards}
\bibfield{author}{\bibinfo{person}{Benwen Zhang}, \bibinfo{person}{Emily Hill}, {and} \bibinfo{person}{James Clause}.} \bibinfo{year}{2016}\natexlab{}.
\newblock \showarticletitle{Towards automatically generating descriptive names for unit tests}. In \bibinfo{booktitle}{\emph{Proceedings of the 31st IEEE/ACM International Conference on Automated Software Engineering}}. \bibinfo{pages}{625--636}.
\newblock


\bibitem[Zhang et~al\mbox{.}(2021)]%
        {zhang2021screen}
\bibfield{author}{\bibinfo{person}{Xiaoyi Zhang}, \bibinfo{person}{Lilian de Greef}, \bibinfo{person}{Amanda Swearngin}, \bibinfo{person}{Samuel White}, \bibinfo{person}{Kyle Murray}, \bibinfo{person}{Lisa Yu}, \bibinfo{person}{Qi Shan}, \bibinfo{person}{Jeffrey Nichols}, \bibinfo{person}{Jason Wu}, \bibinfo{person}{Chris Fleizach}, {et~al\mbox{.}}} \bibinfo{year}{2021}\natexlab{}.
\newblock \showarticletitle{Screen recognition: Creating accessibility metadata for mobile applications from pixels}. In \bibinfo{booktitle}{\emph{Proceedings of the CHI Conference on Human Factors in Computing Systems}}. \bibinfo{pages}{1--15}.
\newblock


\bibitem[Zheng et~al\mbox{.}(2017)]%
        {zheng2017code}
\bibfield{author}{\bibinfo{person}{Wenhao Zheng}, \bibinfo{person}{Hong-Yu Zhou}, \bibinfo{person}{Ming Li}, {and} \bibinfo{person}{Jianxin Wu}.} \bibinfo{year}{2017}\natexlab{}.
\newblock \showarticletitle{Code attention: Translating code to comments by exploiting domain features}.
\newblock \bibinfo{journal}{\emph{arXiv preprint arXiv:1709.07642}} (\bibinfo{year}{2017}).
\newblock


\bibitem[Zhou et~al\mbox{.}(2023)]%
        {zhou2023towards}
\bibfield{author}{\bibinfo{person}{Ziyi Zhou}, \bibinfo{person}{Huiqun Yu}, \bibinfo{person}{Guisheng Fan}, \bibinfo{person}{Zijie Huang}, {and} \bibinfo{person}{Kang Yang}.} \bibinfo{year}{2023}\natexlab{}.
\newblock \showarticletitle{Towards Retrieval-Based Neural Code Summarization: A Meta-Learning Approach}.
\newblock \bibinfo{journal}{\emph{IEEE Transactions on Software Engineering}} (\bibinfo{year}{2023}).
\newblock


\bibitem[Zhou et~al\mbox{.}(2022)]%
        {zhou2022summarizing}
\bibfield{author}{\bibinfo{person}{Ziyi Zhou}, \bibinfo{person}{Huiqun Yu}, \bibinfo{person}{Guisheng Fan}, \bibinfo{person}{Zijie Huang}, {and} \bibinfo{person}{Xingguang Yang}.} \bibinfo{year}{2022}\natexlab{}.
\newblock \showarticletitle{Summarizing source code with hierarchical code representation}.
\newblock \bibinfo{journal}{\emph{Information and Software Technology}}  \bibinfo{volume}{143} (\bibinfo{year}{2022}), \bibinfo{pages}{106761}.
\newblock


\end{thebibliography}

\end{document}